\documentclass[twocolumn]{aastex62}
\hypersetup{linkcolor=magenta, citecolor=cyan, filecolor=green, urlcolor=blue}

\def\arraystretch{0.8}  

\newcommand{\LCDM}{$\Lambda$CDM}

\newcommand{\um}{$\mu$m}

\newcommand{\sig}{$\sigma$}
\newcommand{\Lya}{Lyman-$\alpha$}

\newcommand{\lam}{$\lambda$}

\newcommand{\tentothe}[1]{$10^{#1}$}

\newcommand{\e}[1]{$\times 10^{#1}$}

\newcommand{\sinv}{s$^{-1}$}

\newcommand{\supa}{$^{\rm a}$}
\newcommand{\supb}{$^{\rm b}$}
\newcommand{\supc}{$^{\rm c}$}
\newcommand{\supd}{$^{\rm d}$}
\newcommand{\supe}{$^{\rm e}$}
\newcommand{\supf}{$^{\rm f}$}
\newcommand{\supg}{$^{\rm g}$}

\newcommand{\squared}{$^2$}

\newcommand{\sqarcmin}{arcmin\squared}

\newcommand{\per}{$^{-1}$}
\newcommand{\inv}{\per}

\newcommand{\Lstar}{$L^*$}

\newcommand{\Msun}{$M_\odot$}

\newcommand{\Hb}{H$\beta$}

\newcommand{\II}{\,{\sc ii}}
\newcommand{\III}{\,{\sc iii}}
\newcommand{\IV}{\,{\sc iv}}
\newcommand{\V}{\,{\sc v}}

\newcommand{\HI}{H\,{\sc i}}
\newcommand{\HII}{H\,{\sc ii}}

\newcommand{\CII}{C\,{\sc ii}}
\newcommand{\CIII}{C\,{\sc iii}}
\newcommand{\CIV}{C\,{\sc iv}}

\newcommand{\OIII}{O\,{\sc iii}}

\newcommand{\Om}{\Omega_m}
\newcommand{\OL}{\Omega_\Lambda}

\newcommand{\etal}{et al.}

\newcommand{\HST}{{\em HST}}
\newcommand{\SST}{{\em SST}}
\newcommand{\Hubble}{{\em Hubble}}
\newcommand{\Spitzer}{{\em Spitzer}}
\newcommand{\Chandra}{{\em Chandra}}
\newcommand{\XMM}{{\em XMM-Newton}}
\newcommand{\JWST}{{\em JWST}}
\newcommand{\Planck}{{\em Planck}}

\newcommand{\Bradac}{{Brada\v{c}}}

\newcommand{\citepeg}[1]{\citep[e.g.,][]{#1}}

\received{---}
\revised{---}
\accepted{---}
\submitjournal{ApJ}

\shorttitle{RELICS: Reionization Lensing Cluster Survey}
\shortauthors{Coe \etal\ 2019}

\begin{document}

\title{\bf \large RELICS: Reionization Lensing Cluster Survey}

\correspondingauthor{Dan Coe}
\email{DCoe@STScI.edu}

\author[0000-0001-7410-7669]{Dan Coe} 
\affil{Space Telescope Science Institute, 3700 San Martin Drive, Baltimore, MD 21218, USA}

\author[0000-0002-7453-7279]{Brett Salmon} 
\affil{Space Telescope Science Institute, 3700 San Martin Drive, Baltimore, MD 21218, USA}

\author[0000-0001-5984-0395]{Maru\v{s}a Brada\v{c}} 
\affil{Department of Physics, University of California, Davis, CA 95616, USA}

\author[0000-0002-7908-9284]{Larry D. Bradley} 
\affil{Space Telescope Science Institute, 3700 San Martin Drive, Baltimore, MD 21218, USA}

\author[0000-0002-7559-0864]{Keren Sharon} 
\affiliation{Department of Astronomy, University of Michigan, 1085 South University Ave, Ann Arbor, MI 48109, USA}

\author[0000-0002-0350-4488]{Adi Zitrin} 
\affil{Physics Department, Ben-Gurion University of the Negev, P.O. Box 653, Beer-Sheva 84105, Israel}

\author[0000-0003-3108-9039]{Ana Acebron} 
\affil{Physics Department, Ben-Gurion University of the Negev, P.O. Box 653, Beer-Sheva 84105, Israel}

\author[0000-0002-8261-9098]{Catherine Cerny} 
\affil{Astronomy Department and Institute for Astrophysical Research, Boston University, 725 Commonwealth Ave., Boston, MA 02215, USA}

\author{Nath\'alia Cibirka} 
\affil{Physics Department, Ben-Gurion University of the Negev, P.O. Box 653, Beer-Sheva 84105, Israel}

\author[0000-0002-6338-7295]{Victoria Strait} 
\affil{Department of Physics, University of California, Davis, CA 95616, USA}

\author[0000-0003-3653-3741]{Rachel Paterno-Mahler} 
\affil{Department of Astronomy, University of Michigan, 1085 South University Ave, Ann Arbor, MI 48109, USA}

\author[0000-0003-3266-2001]{Guillaume Mahler} 
\affil{Department of Astronomy, University of Michigan, 1085 South University Ave, Ann Arbor, MI 48109, USA}

\author[0000-0001-9364-5577]{Roberto J. Avila} 
\affil{Space Telescope Science Institute, 3700 San Martin Drive, Baltimore, MD 21218, USA}

\author{Sara Ogaz} 
\affil{Space Telescope Science Institute, 3700 San Martin Drive, Baltimore, MD 21218, USA}

\author{Kuang-Han Huang} 
\affil{Department of Physics, University of California, Davis, CA 95616, USA}

\author{Debora Pelliccia} 
\affil{Department of Physics, University of California, Davis, CA 95616, USA}
\affil{Department of Physics and Astronomy, University of California, Riverside, CA 92521, USA}

\author{Daniel P. Stark} 
\affil{Department of Astronomy, Steward Observatory, University of Arizona, 933 North Cherry Avenue, Tucson, AZ, 85721, USA}

\author{Ramesh Mainali} 
\affil{Department of Astronomy, Steward Observatory, University of Arizona, 933 North Cherry Avenue, Tucson, AZ, 85721, USA}

\author{Pascal A. Oesch} 
\affil{Department of Astronomy, University of Geneva, Chemin des Maillettes 51, 1290 Versoix, Switzerland}

\author{Michele Trenti} 
\affil{School of Physics, University of Melbourne, VIC 3010, Australia}
\affil{ARC Centre of Excellence for All Sky Astrophysics in 3 Dimensions (ASTRO 3D), VIC 2010, Australia}

\author[0000-0002-3772-0330]{Daniela Carrasco} 
\affil{School of Physics, University of Melbourne, VIC 3010, Australia}

\author[0000-0003-0248-6123]{William A. Dawson} 
\affil{Lawrence Livermore National Laboratory, P.O. Box 808 L-210, Livermore, CA, 94551, USA}

\author{Steven A. Rodney} 
\affil{Department of Physics and Astronomy, University of South Carolina, 712 Main St., Columbia, SC 29208, USA}

\author[0000-0002-7756-4440]{Louis-Gregory Strolger} 
\affil{Space Telescope Science Institute, 3700 San Martin Drive, Baltimore, MD 21218, USA}

\author{Adam G. Riess} 
\affil{Space Telescope Science Institute, 3700 San Martin Drive, Baltimore, MD 21218, USA}

\author{Christine Jones} 
\affil{Harvard-Smithsonian Center for Astrophysics, 60 Garden Street, Cambridge, MA 02138, USA}

\author[0000-0003-1625-8009]{Brenda L. Frye} 
\affil{Department of Astronomy, Steward Observatory, University of Arizona, 933 North Cherry Avenue, Tucson, AZ, 85721, USA}

\author{Nicole G. Czakon} 
\affil{Academia Sinica Institute of Astronomy and Astrophysics (ASIAA), No.~1, Section 4, Roosevelt Road, Taipei 10617, Taiwan}

\author[0000-0002-7196-4822]{Keiichi Umetsu} 
\affil{Academia Sinica Institute of Astronomy and Astrophysics (ASIAA), No.~1, Section 4, Roosevelt Road, Taipei 10617, Taiwan}

\author[0000-0003-0980-1499]{Benedetta Vulcani} 
\affil{INAF-Osservatorio Astronomico di Padova, Vicolo Dell'osservatorio 5, 35122 Padova Italy}

\author{Or Graur}
\affil{Harvard-Smithsonian Center for Astrophysics, 60 Garden Street, Cambridge, MA 02138, USA}
\affil{Department of Astrophysics, American Museum of Natural History, Central Park West and 79th Street, New York, NY 10024-5192, USA}
\affil{NSF Astronomy and Astrophysics Postdoctoral Fellow}

\author[0000-0001-8738-6011]{Saurabh W. Jha}
\affil{Department of Physics and Astronomy, Rutgers, The State University of New Jersey, 136 Frelinghuysen Road, Piscataway, NJ 08854, USA}

\author{Melissa L. Graham}
\affil{Department of Astronomy, University of Washington, Box 351580, U.W., Seattle, WA 98195-1580, USA}

\author{Alberto Molino}
\affil{Universidade de S$\tilde a$o Paulo, Instituto de Astronomia, Geof\'isica e Ci\^encias Atmosf\'ericas, Rua do Mat$\tilde a$o 1226, S$\tilde a$o Paulo, Brazil}
\affil{Instituto de Astrof\'isica de Andaluc\'ia (IAA-CSIC), E-18080 Granada, Spain}

\author{Mario Nonino}
\affil{INAF - Osservatorio Astronomico di Trieste, Via Tiepolo 11, I-34131 Trieste, Italy}

\author[0000-0002-4571-2306]{Jens Hjorth}
\affil{DARK, Niels Bohr Institute, University of Copenhagen, Lyngbyvej 2, DK-2100 Copenhagen, Denmark}

\author[0000-0001-9058-3892]{Jonatan Selsing}
\affil{Cosmic Dawn Center (DAWN)}
\affil{Niels Bohr Institute, University of Copenhagen, Lyngbyvej 2, DK-2100, Copenhagen, Denmark}

\author[0000-0001-8415-7547]{Lise Christensen}
\affil{DARK, Niels Bohr Institute, University of Copenhagen, Lyngbyvej 2, DK-2100 Copenhagen, Denmark}

\author{Shotaro Kikuchihara} 
\affil{Institute for Cosmic Ray Research, The University of Tokyo, 5-1-5 Kashiwanoha, Kashiwa, Chiba 277-8582, Japan}
\affil{Department of Astronomy, Graduate School of Science, The University of Tokyo, 7-3-1 Hongo, Bunkyo, Tokyo, 113-0033, Japan}

\author[0000-0002-1049-6658]{Masami Ouchi} 
\affil{Institute for Cosmic Ray Research, The University of Tokyo, 5-1-5 Kashiwanoha, Kashiwa, Chiba 277-8582, Japan}
\affil{Kavli Institute for the Physics and Mathematics of the Universe (Kavli IPMU, WPI), University of Tokyo, Kashiwa, Chiba 277-8583, Japan}

\author{Masamune Oguri} 
\affil{Research Center for the Early Universe, The University of Tokyo, 7-3-1 Hongo, Bunkyo-ku, Tokyo 113-0033, Japan}
\affil{Department of Physics, The University of Tokyo, 7-3-1 Hongo, Bunkyo-ku, Tokyo 113-0033, Japan}
\affil{Kavli Institute for the Physics and Mathematics of the Universe (Kavli IPMU, WPI), University of Tokyo, Kashiwa, Chiba 277-8583, Japan}

\author{Brian Welch} 
\affil{Department of Physics and Astronomy, The Johns Hopkins University, 3400 North Charles Street, Baltimore, MD 21218, USA}

\author[0000-0002-1428-7036]{Brian C. Lemaux} 
\affil{Department of Physics, University of California, Davis, CA 95616, USA}

\author[0000-0002-8144-9285]{Felipe Andrade-Santos} 
\affil{Harvard-Smithsonian Center for Astrophysics, 60 Garden Street, Cambridge, MA 02138, USA}

\author{Austin T. Hoag} 
\affil{Department of Physics, University of California, Davis, CA 95616, USA}

\author[0000-0002-8829-5303]{Traci L. Johnson} 
\affil{Department of Astronomy, University of Michigan, 1085 South University Drive, Ann Arbor, MI 48109, USA}


\author{Avery Peterson} 
\affil{Department of Astronomy, University of Michigan, 1085 South University Drive, Ann Arbor, MI 48109, USA}

\author{Matthew Past} 
\affil{Department of Astronomy, University of Michigan, 1085 South University Drive, Ann Arbor, MI 48109, USA}


\author{Carter Fox} 
\affil{Department of Astronomy, University of Michigan, 1085 South University Ave, Ann Arbor, MI 48109, USA}

\author{Irene Agulli} 
\affil{Physics Department, Ben-Gurion University of the Negev, P.O. Box 653, Beer-Sheva 84105, Israel}

\author{Rachael Livermore} 
\affil{School of Physics, University of Melbourne, VIC 3010, Australia}
\affil{ARC Centre of Excellence for All Sky Astrophysics in 3 Dimensions (ASTRO 3D), VIC 2010, Australia}

\author{Russell E. Ryan} 
\affil{Space Telescope Science Institute, 3700 San Martin Drive, Baltimore, MD 21218, USA}

\author{Daniel Lam} 
\affil{Leiden Observatory, Leiden University, NL-2300 RA Leiden, The Netherlands}

\author{Irene Sendra-Server} 
\affil{Department of Theoretical Physics, University of the Basque Country UPV/EHU, E-48080 Bilbao, Spain}

\author[0000-0003-3631-7176]{Sune Toft} 
\affil{Cosmic Dawn Center (DAWN)}
\affil{Niels Bohr Institute, University of Copenhagen, Lyngbyvej 2, DK-2100, Copenhagen, Denmark}

\author{Lorenzo Lovisari} 
\affil{Harvard-Smithsonian Center for Astrophysics, 60 Garden Street, Cambridge, MA 02138, USA}

\author{Yuanyuan Su} 
\affil{Harvard-Smithsonian Center for Astrophysics, 60 Garden Street, Cambridge, MA 02138, USA}








\renewcommand{\~}{$\sim$}

\begin{abstract}
Large surveys of galaxy clusters with the \Hubble\ and \Spitzer\ {\em Space Telescopes},
including CLASH and the Frontier Fields,
have demonstrated the power of strong gravitational lensing
to efficiently deliver large samples of high-redshift galaxies.
We extend this strategy through a wider, shallower survey named RELICS,
the Reionization Lensing Cluster Survey.
This survey, described here, was designed primarily to deliver the best and brightest high-redshift candidates 
from the first billion years after the Big Bang.
%
RELICS observed 41 massive galaxy clusters 
with \Hubble\ and \Spitzer\ at 0.4--1.7\um\ and 3.0--5.0\um, respectively.
We selected 21 clusters based on \Planck\ PSZ2 mass estimates
and the other 20 based on observed or inferred lensing strength.
Our 188-orbit \Hubble\ Treasury Program obtained the first high-resolution near-infrared images of these clusters
to efficiently search for lensed high-redshift galaxies.
We observed 46 WFC3/IR pointings (\~200 \sqarcmin) with two orbits divided among four filters (F105W, F125W, F140W, and F160W)
and ACS imaging as needed to achieve single-orbit depth in each of three filters (F435W, F606W, and F814W).
As previously reported by Salmon et al., 
we discovered 322 $z \sim 6-10$ candidates, including the brightest known at $z \sim 6$,
and the most spatially-resolved distant lensed arc known at $z \sim 10$.
\Spitzer\ IRAC imaging (945 hours awarded, plus 100 archival) 
has crucially enabled us to distinguish $z \sim 10$ candidates from $z \sim 2$ interlopers.
For each cluster, two \HST\ observing epochs were staggered by about a month, enabling us to discover 11 supernovae, 
including 3 lensed supernovae, which we followed up with 20 orbits from our program.
We delivered reduced \HST\ images and catalogs of all clusters to the public via MAST
and reduced \Spitzer\ images via IRSA.
We have also begun delivering lens models of all clusters,
to be completed before the  \JWST\ GO Cycle 1 call for proposals.
\end{abstract}


\keywords{early universe --
galaxies: high redshift --
cosmology: dark ages, reionization, first stars --
galaxies: evolution --
gravitational lensing: strong --
galaxies: clusters
}


\section{Introduction}
\label{sec:intro}

Gravitational lensing magnification by massive galaxy clusters 
has a long history of helping astronomers discover the most distant galaxies known
with the \Hubble\ and \Spitzer\ Space Telescopes
(see \citealt{KneibNatarajan11} \S5.8 for a review).
Twenty years ago, a $z = 4.92$ galaxy lensed by the cluster MS1358+62 was the most distant known \citep{Franx97}.
Ten years ago, that record belonged to A1689-zD1 at $z \sim 7.5$ \citep{Bradley08, Watson15}.
More recently, the Cluster Lensing And Supernova survey with Hubble (CLASH; \citealt{CLASH})
yielded the triply-imaged $z \sim 10.8$ candidate MACS0647-JD \citep{Coe13, Pirzkal15, Chan16}.
This redshift has been surpassed only by the grism measurement of $z = 11.1$ \citep{Oesch16}
for GN-z11 discovered by \cite{Oesch14a}
in blank-field imaging by the CANDELS program \citep{Grogin11,Koekemoer11}.
Both $z \sim 11$ galaxies are similarly bright at 1.6\um\ (F160W AB mag 25.9),
making them excellent targets for follow-up study with
the {\em James Webb Space Telescope (JWST)}.

For a given observing strategy,
lensed fields generally yield significantly more high-redshift galaxies than blank fields, 
especially at the highest redshifts and in relatively shallow imaging
\citep[e.g.,][]{Coe15}.
Lensing does sacrifice high-redshift search volume due to the magnification as well as the foreground cluster light.
But lensing more than compensates for the lost volume
by magnifying many more faint galaxies into view
whenever luminosity function number counts are steeper than $\phi \propto L^{-2}$ \citep{Broadhurst95}.
This is especially true brightward of \Lstar, for example, AB mag \~28 at $z \sim 8$ \citep{Bradley12b,Finkelstein16}.

CLASH demonstrated this lensing advantage,
yielding significantly more galaxies at $z \sim 6 - 8$ than blank field surveys \citep{Bradley14}.
CLASH obtained 20-orbit \HST\ imaging for each of 25 clusters in 16 filters,
including F160W observed to AB mag 27.5 (5\sig\ depth; \citealt{CLASH}; \citealt{Molino17}).
The first 18 clusters yielded 262 candidates at $z \sim 6 - 8$ \citep{Bradley14},
plus a few at $z \sim 9 - 11$ \citep{Zheng12N,Coe13,Bouwens14b}.

The Frontier Fields \citep{Lotz16} obtained deeper 140-orbit \HST\ imaging 
for each of 6 clusters and 6 blank parallel fields
in 7 filters, including F160W observed to AB mag 28.7.
At these depths, $\phi \propto L^{-2}$ for high-redshift galaxies 
(the luminosity function faint end slope $\alpha \sim -2$ roughly at $z \sim 6 - 8$),
so the lensed and blank field high-z counts were predicted to be similar \citep{Coe15}.
Indeed, the Frontier Fields yielded similar total numbers of lensed and blank field high-redshift galaxies, 
altogether 453 at $z \sim 6 - 9$ from one analysis \citep{Kawamata18},
plus a few $z \sim 10$ candidates \citep{Zitrin14,Infante15,McLeod16,Oesch18}.
Due to their magnifications, the lensed galaxies are intrinsically fainter than the blank field galaxies.
By combining deep \Hubble\ imaging with the power of gravitational lensing,
the Frontier Fields revealed the faintest galaxies yet known,
including some of those likely responsible for reionization at $z \sim 6$ \citep{Livermore16,Atek18}.

While CLASH and the Frontier Fields yielded many high-redshift candidates,
neither survey was optimized to deliver high-redshift candidates
observed brightly enough for detailed follow-up study with current and future observatories. 
Detailed studies are required to determine galaxies' ages, stellar masses, compositions, and ionizing strengths
(see \citealt{Stark16} for a review).
It is imperative to use existing facilities to discover the best targets for study in the early universe,
ideally before \JWST\ Cycle 1.
Gravitational lensing offers the most efficient route to do this.
The Planck all-sky survey delivered the PSZ2 catalog of $>$1,000 massive galaxy clusters \citep{Planck15XXVII}
detected via their Sunyaev-Zel'dovich (SZ) effect \citep{SunyaevZeldovich70} on the Cosmic Microwave Background (CMB).
%
By searching the PSZ2 catalog,
we found that many massive clusters (including presumably excellent lenses)
lacked the combination of \HST\ Advanced Camera for Surveys (ACS)
and Wide Field Camera 3 infrared (WFC3/IR) 
imaging required to discover high-redshift candidates at $z > 6$.

RELICS embarked on an efficient survey with \Hubble\ and \Spitzer\ 
to discover the best and brightest high-redshift galaxies \citep{Coe18}.
RELICS obtained 5-orbit \HST\ imaging of 41 clusters in 7 filters 
(the same filters used by the Frontier Fields), including F160W to AB mag 26.5.
Our relatively shallow imaging covered more area than CLASH or the Frontier Fields, 
yielding high-redshift candidates that are brighter,
either intrinsically and/or due to lensing magnification.
See Table \ref{tab:surveys} for a summary comparison of CLASH, the Frontier Fields, and RELICS.
More recently, a new \HST\ program BUFFALO (PIs Steinhardt \& Jauzac; GO 15117)
has begun extending the Frontier Fields to the wider area covered by deep \Spitzer\ imaging.


Also notable and inspiring for this project are the \HST\ Snapshot programs observing
galaxy clusters discovered by the MAssive Cluster Survey (MACS; \citealt{Ebeling01}).
Shallower \HST\ imaging ($\sim 1$ orbit per cluster) has been obtained for 86 clusters to date \citep{ReppEbeling18}.
For 29 of these clusters, 4-band imaging, including F140W to AB mag 26.6, was completed,
yielding 20 candidates at $z \sim 7 - 9$ \citep{Repp16}.
Nine MACS clusters are included in RELICS.
Prior to RELICS, they had archival \HST\ ACS and/or WFPC2 imaging, but not WFC3/IR or NICMOS.

Meanwhile, large \Spitzer\ programs such as SURFS UP \citep{Bradac14, Ryan14, Huang16} 
and the Frontier Fields
have also surveyed many galaxy clusters, 
helping to identify high-redshift galaxies and study their properties \citepeg{Hoag19}.
The Grism Lens-Amplified Survey from Space (GLASS; \citealt{Treu15})
delivered spectroscopic properties and high-redshift confirmations \citepeg{Schmidt16,Schmidt17}.
The high-redshift searches in lensing programs complement searches in blank field surveys such as the UDF, CANDELS, and BoRG,
constraining luminosity functions from $z \sim 4 - 10$ \citep{Bouwens15,Finkelstein16,Oesch18,Morishita18}.

Below we discuss the RELICS science drivers (\S\ref{sec:sci}),
galaxy cluster targets (\S\ref{sec:clus}),
observations with \Hubble, \Spitzer, and other observatories (\S\ref{sec:obs}),
the image reductions and catalogs (\S\ref{sec:reductions}),
and results to date (\S\ref{sec:results}),
followed by a summary (\S\ref{sec:summary}).

We use the AB magnitude system,
$m_{\rm AB} = 31.4 - 2.5 \log(f_\nu / {\rm nJy})$
\citep{Oke74,OkeGunn83}.
%
Where needed, 
we assume a flat concordance \LCDM\ cosmology
with $h = 0.7$, $\Om = 0.3$, and $\OL = 0.7$,
where $H_0 = 100 \, h$ km s\inv\ Mpc\inv.
Galaxy cluster masses are given as $M_{500}$,
the mass enclosed within $r_{500}$,
within which the average density is 500 times the critical density of the universe at that redshift \citep[e.g.,][]{Coe10DMprofiles}.
These masses $M_{500}$ are less than the total virial cluster masses measured at larger radii.


\begin{deluxetable}{lccc}
\tablecaption{\label{tab:surveys}Recent Large \HST\ Cluster Lensing Surveys}
\tablewidth{\columnwidth}
\tablehead{
\colhead{}&
\colhead{CLASH}&
\colhead{Frontier Fields}&
\colhead{RELICS}
}
\startdata
Clusters & 25 & 6 & 41\\
\HST\ orbits per cluster & 20\supa & 140 & 5\supa\\
Total \HST\ orbits & 524 & 840 & 188\\
Supernova orbits\supb & 50 & 0 & 20\\
\HST\ Filters & 16 & 7 & 7\\
Depth in F160W\supc & 27.5 & 28.7 & 26.5\\
\HST\ Cycle Numbers & 18 -- 20 & 21 -- 23 & 23\\
\HST\ Begin & Nov.'10 & Oct.'13 & Oct.'15\\
\HST\ End & Jul.'13 & Sept.'16 & Apr.'17
\enddata
\tablenotetext{a}{Depth including archival \HST\ imaging}
\tablenotetext{b}{\HST\ orbits allocated specifically for supernova follow-up}
\tablenotetext{c}{AB mag 5\sig\ depth for point sources}
\end{deluxetable}

\begin{figure}
\centerline{
\includegraphics[width = \columnwidth]{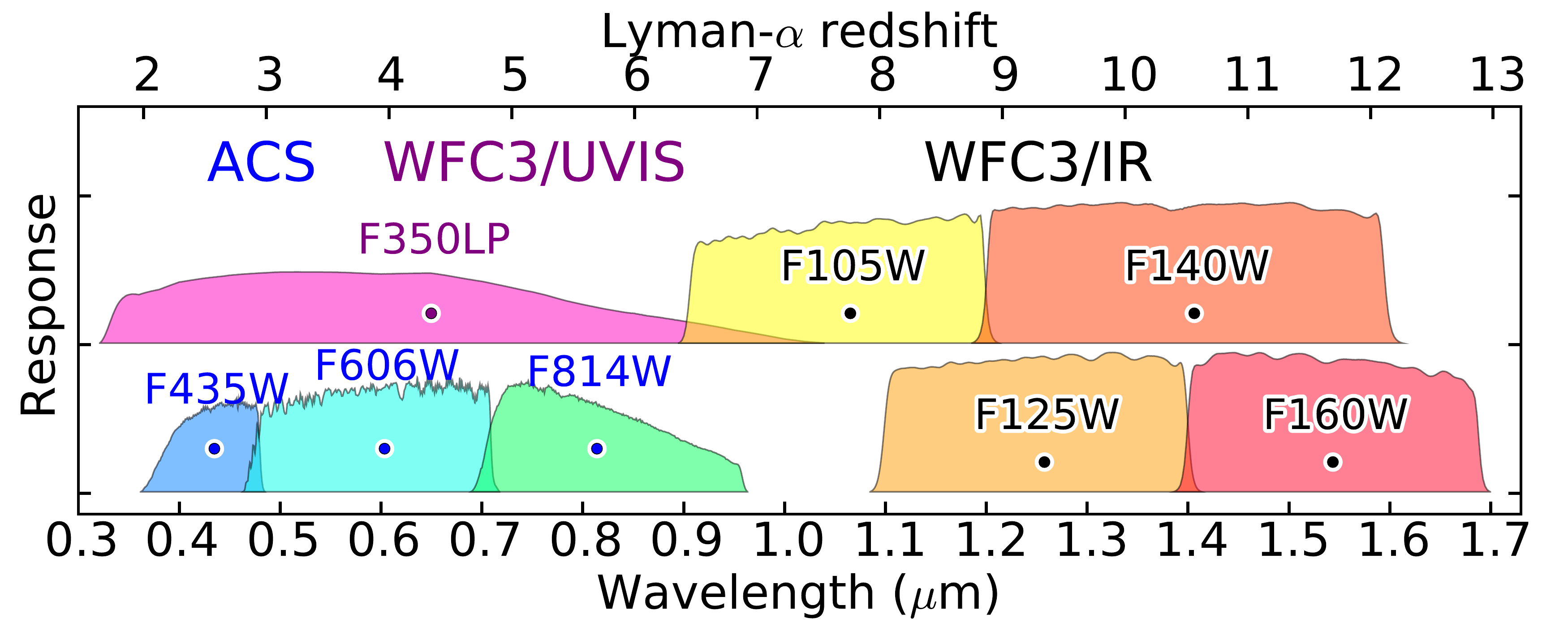}
}
\caption{
\label{fig:filters}
RELICS clusters were observed with the seven \Hubble\ ACS and WFC3/IR filters shown here.
RELICS parallels were observed with the one WFC3/UVIS and four WFC3/IR filters shown.
Response curves are plotted versus wavelength ($\lambda$)
with the corresponding \Lya\ redshift ($z$) given along the top axis ($\lambda = 0.1216\mu$m $(1+z)$).
F350LP, F105W, and F140W are offset vertically for clarity.
Black dots mark the effective ``pivot'' wavelengths \citep{TokunagaVacca05} of the filters.
}
\end{figure}


\section{Science}
\label{sec:sci}

RELICS was primarily designed and optimized to search for brightly lensed high-redshift galaxies in the epoch of reionization.
Ancillary science enabled by RELICS includes
supernova searches, cluster mass measurements, and limits on the dark matter particle cross section.
We discuss each of these science goals in turn.

\subsection{High-Redshift Galaxies}
\label{sec:highz}

Hubble's WFC3 has revealed \~2,000 $z \sim 6-11$ candidates from the Universe's first billion years, 
but only a small fraction have been bright enough ($H<25.5$)
for detailed follow-up study and spectroscopic confirmation \citepeg{Bouwens15,Salmon17}.
RELICS was designed primarily to deliver 
1) brightly observed high-redshift candidates, amenable to more detailed studies; and
2) a large sample of high-redshift candidates to improve luminosity function constraints.
This combination is required to improve our understanding of galaxies and reionization in the first billion years.

\subsubsection{Spectroscopic Studies of Brightly Observed High-Redshift Candidates}

Spectroscopic redshift confirmations based on \Lya\ detections have proven increasingly difficult beyond $z > 6$ 
\citep{Stark10, Bradac12, Schenker14, Schmidt16, Hoag19, Mason19}, 
likely due primarily to absorption by patchy neutral hydrogen before reionization was complete 
\citep{Treu13, Tilvi14, Mason18a}.
A higher success rate was achieved with four luminous $z \sim 7 - 9$ galaxies 
inferred to have significant [\OIII]+\Hb\ equivalent width (EW) based on \Spitzer\ photometry \citep{Roberts-Borsani16}.
All four yielded \Lya\ detections as distant as $z = 8.68$ \citep{Oesch15a,Zitrin15,Stark17}.
This suggests that luminous galaxies with high ionization parameters carve out ionized gas bubbles
allowing \Lya\ to stream free \citep{Stark17}.

Alternatives to \Lya\ redshift confirmations include fainter, slightly redder UV metal lines 
N\V\lam1243\AA, C\IV\lam1549\AA, He\II\lam1640\AA, O\III]\lam1663\AA, and C\III]\lam1908\AA\ 
\citep{Stark14} 
and sub-mm lines
including [C\II] 158\um\ and [O\III] 88\um\ 
\citep{Inoue14}.
Studying these lines also yields more information about the physical properties of the galaxies, including their ionizing strength.

The UV metal line \CIII] has been detected in several $z > 6$ galaxies as distant as $z = 7.73$ 
\citep{Stark15,Stark17,Laporte17a,Mainali18}.
At $z \sim 2$, \cite{Rigby15} found that lower metallicity galaxies exhibit stronger \CIII],
likely explaining why \CIII] appears far more often at $z \sim 6$ than at lower redshifts
\citep[see also][]{Senchyna17,Du17,Le-Fevre17}.
C\III] detections may be powered by low metallicity massive stars \citep{Stark15,Stark17}
or may require AGNs \citep{Nakajima18a}.
Still lower metallicity stars ($< 0.05 Z_\odot$) produce higher ionization potentials yielding \CIV. 
All three known $z > 6$ \CIV\ detections are lensed \citep{Stark15, Mainali17, Hoag17} 
and thus less massive and presumably lower metallicity 
than average $z > 6$ galaxies from current surveys.

Larger samples of UV metal line observations including RELICS galaxies 
will enable us to quantify the prevalence of these intense ionizing sources in the reionization epoch,
both directly and by extrapolating to larger samples with similar observed properties 
(luminosity, rest-frame UV slope $\beta$, and \Lya). 
The observed equivalent widths will also be very informative to 
future \JWST\ surveys planning to observe these spectral features.

A new window for studying high-redshift galaxies has been opened by
the Atacama Large Millimeter/submillimeter Array (ALMA) in the south
and the Plateau de Bure interferometer (PdBI) in the north,
later upgraded and renamed the Northern Extended Millimeter Array (NOEMA).
These telescopes (primarily ALMA)
have detected high-redshift galaxies by their infrared continuum dust emission \citep{Capak15}
as well as spectral lines [\CII] 158\um\ and [\OIII] 88\um, which trace star formation \citep{DeLooze14}.
Continuum observations show that $z \sim 5 - 6$ galaxies have at least an order of magnitude less dust
than local starbursts with similar rest-frame UV colors \citep{Capak15}.
Dust is often not detected at these redshifts, but at least a few galaxies have been found to be dusty and thus evolved
as early as $z = 7.15$ \citep{Hashimoto18b}, $z = 7.5$ \citep{Watson15}, $z = 8.312$ \citep{Tamura18}, and $z = 8.38$ \citep{Laporte17a}.

[\CII] 158\um\ is one of the brightest lines in local galaxies \citep{Malhotra97,Brauher08} 
and the strongest observed ISM cooling line in $z \sim 1 - 2$ galaxies \citep{Stacey10}. 
To date, [\CII] has been detected in 22 galaxies between $5.1533 \leq z \leq 7.1453$
(\citealt{Carniani18b} and references therein; \citealt{Hashimoto18b}),
including two spatially resolved galaxies at $z = 6.81$ and 6.85 displaying disk-like rotation \citep{Smit18}.

At higher redshifts, \cite{Inoue14} predicted that [\OIII] 88\um\ would yield more detections.
While [\CII] is associated with neutral \HI\ gas in photodissociation regions (PDRs),
[\OIII] is associated with ionized \HII\ gas, 
more prevalent in higher redshift, lower metallicity galaxies with higher ionization states \citep{Harikane18b}.
As predicted, ALMA's highest-redshift spectroscopic confirmations have come from [\OIII],
and the six $z > 6$ galaxies targeted to date have all yielded [\OIII] detections at
$z = 6.900$ \citep{Marrone18};
$z = 7.107$ \citep{Carniani17};
$z = 7.212$ \citep{Inoue16};
$z = 8.312$ \citep{Tamura18};
$z = 8.382$ \citep{Laporte17c};
and $z = 9.11$ \citep{Hashimoto18a}.
The two highest redshift detections are lensed galaxies.

Based on this previous work, 
we expect RELICS to deliver ALMA confirmations and science at $z \sim 6 - 10$
(see \S\ref{sec:highzresults}).


\subsubsection{Luminosity Functions of Galaxies at the Epoch of Reionization}

Improving constraints on the $z \sim 9$ luminosity function 
(and the evolution from $z \sim 10$ to 6)
is another primary science goal of RELICS.
This is required to determine the numbers of faint galaxies available to reionize the universe.


Planck constrained the reionization history 
by measuring the column density of free electrons to the CMB,
or the Thomson scattering optical depth $\tau = 0.058 \pm 0.012$ 
\citep{Planck18VI}.
This value, lower than previous estimates, implies a later reionization
halfway complete by $z \sim 8$, before fully completing by $z \sim 6$.
It also means reionization can be achieved 
by galaxies producing less Lyman-continuum (LyC) flux and with lower escape fractions $f_{esc}$.
Direct measurements of these $f_{esc}$ values have been obtained recently for \~40 galaxies with detected LyC leakage,
including half extending from the local universe \citepeg{Leitherer16} to $z \sim 0.3$ \citepeg{Izotov18b},
with the other half at $z \sim 2 - 4$ \citepeg{Shapley16,Rivera-Thorsen17b,Vanzella18}.
Assuming these measurements hold at higher redshifts,
and given high-redshift luminosity functions,
low-mass galaxies could have produced most of the flux required to reionize the universe 
\citepeg{Robertson15,Madau17,Ishigaki18,Mason18a,Finkelstein19}.
There may also have been significant contributions from low-luminosity AGN jets at $z \gtrsim 6$ \citep{Bosch-Ramon18}.

At the highest redshifts ($z \sim 9-12$),
\Hubble\ and \Spitzer\ imaging programs have yielded fewer candidates than expected.
This has left luminosity functions highly uncertain at these redshifts,
while also hinting at accelerated evolution in the first 600 Myr 
\citep{Bouwens12b,Oesch18}.
Luminosity functions are fairly well constrained at $z \sim 4 - 8$ \citep{Bouwens15,Finkelstein16,Kawamata18}.
Only recently were constraints placed on all three parameters ($\phi^*$, $M^*$, $\alpha$)
of the $z \sim 9$ luminosity function \citep{Morishita18};
these results are consistent with either accelerated or smoother (non-accelerated) evolution at $z > 8$.
Higher precision measurements, especially with greater leverage at higher redshifts,
are required to better constrain this early evolution rate.

\subsection{Strong Lens Modeling}
\label{sec:lensing}

Robust strong lens modeling is required to estimate magnifications of our lensed galaxies and the survey volume.
The mass in galaxy cluster cores responsible for the lensing is predominantly dark matter.
We must infer the distribution of this dark matter based on 
observed strong lensing that produces multiple images of more distant galaxies.
We often add the assumption that luminous galaxy cluster members are good tracers of their dark matter halos.

To this end, RELICS obtained 7-band \HST\ imaging of each cluster, 
and we are using ground-based telescopes to measure spectroscopic redshifts of strongly lensed galaxies.
High-quality lens modeling generally requires multiband high-resolution \HST\ imaging
to reliably identify multiple images of strongly lensed galaxies
based on their colors and morphologies.
Spectroscopic redshifts are also crucial.
Magnification accuracies improve with greater numbers of strongly lensed images with spectroscopic redshifts
\citep{JohnsonSharon16}.
Recent tests with simulated cluster lensing find that
the best lens models accurately yield magnification estimates of 3 (10) with precisions of 15\% (30\%),
with uncertainty increasing with magnification (\citealt{Meneghetti16}; see also \citealt{Zitrin15CLASH}).
Such precision is encouraging, 
as the vast majority (over 80\%) of lensed high-redshift galaxies observed are magnified by factors of 10 or less \citep{Coe15}.

Magnification uncertainties directly impact measurements of some physical properties such as
luminosity, star formation rate, stellar mass, and size.
However, since lensing is achromatic, other properties derived from galaxy colors are not affected by magnification;
these include redshift, age, metallicity, extinction, and rest-frame UV slope.
Magnifications do affect luminosity functions,
but the uncertainties are mitigated by averaging over many galaxies.
The resulting uncertainties on the total survey volume 
are subdominant compared to the current small number statistics at high redshifts \citep{Coe15}.

In the case of a multiply-imaged high-redshift candidate,
lens modeling can yield geometric support,
distinguishing between low and high-redshift solutions
based on the separation between the observed images
\citepeg{Coe13,Zitrin14,Chan16}.
Lens modeling is required to study delensed (source plane) properties,
including \~100 parsec structures resolved in highly elongated arcs \citepeg{Sharon12,Johnson17}.
Lens modeling is also required to constrain luminosity functions in lensed fields \citepeg{Livermore16}.

RELICS lens modeling is yielding the overall lensing strength for each cluster.
While we expect all RELICS clusters to be excellent lenses,
our analyses will reveal which are truly among the strongest and best to use
for efficient discoveries of the most distant galaxies known in future surveys.
For RELICS results to date, see \S\ref{sec:lensresults}.

\subsection{Galaxy Cluster Masses}
\label{sec:clusters}

Scaling relations linking mass estimates from lensing, X-ray, and SZ studies
require understanding of observational systematics and scale-dependent cluster astrophysics.
Improving the accuracy of these mass scaling relations will be 
key to realizing the full potential of future missions such as eROSITA, 
which anticipates detecting 100,000 clusters / groups out to $z \sim 1.3$.

Constraints on cosmological parameters (primarily $\Om$ and $\sigma_8$)
derived from the Planck CMB \citep{Planck15XIII}
are at odds with those derived from Planck SZ galaxy cluster counts \citep{Planck15XXIV}.
The latter paper calibrated the scaling relation between Planck SZ signal strength and cluster masses
based on analyses of \XMM\ X-ray observations assuming hydrostatic equilibrium (HSE).
One may expect cluster masses to be \~20\% greater due to deviations from HSE, 
including non-thermal pressure support \citepeg{Nagai07, Rasia12}.
But \cite{Planck15XXIV} found that the cosmology results can be reconciled 
only if one assumes that SZ-derived cluster masses underestimate the true $M_{500}$ masses
by $b \approx 42\%$  
(where they reported the mass bias as $1 - b = 0.58 \pm 0.04$).
That is about 2-$\sigma$ greater than the \~20\% initially adopted in the PSZ1 analysis \citep{Planck13XXIX}.



Weak lensing analysis of clusters included in both the Planck analysis
and the Weighing the Giants survey (WtG; \citealt{vonderLinden14b})
shows Planck cluster masses may indeed be underestimated by \~42\% for the most massive clusters ($> 10^{15} M_{\odot}$),
while Planck masses appear to be more accurate for less massive clusters ($\sim 5 \times 10^{14} M_{\odot}$).
Subsequent weak lensing analyses from various surveys (WtG, CCCP, LoCUSS, CLASH, CFHTLenS, RCSLenS, HSC-SSP)
found a range of results, some consistent with WtG including bias increasing with mass \citep{SerenoEttori17}, 
and others more consistent with the original Planck estimate of \~20\% bias
\citep{Hoekstra15,Battaglia16,Smith16,Penna-Lima17,Medezinski18}.
Overall, the tension appears to be somewhat relieved, although not conclusively \citep{Planck18I},
especially after accounting for new Planck measurements of the reionization optical depth \citep{Planck16XLVI,Planck18VI}.

SZ mass estimates are calibrated in part based on X-ray temperatures measured with spectroscopy out to $r_{500}$,
which can be difficult, especially for higher redshift clusters.
X-ray mass calibration at the smaller $r_{2500}$ radii probed by \HST\ strong lensing may be key to improving mass scaling relations.
These smaller radii are also important for ground-based SZ telescopes offering significantly higher resolution than Planck.
\cite{Czakon15} analyzed arcsecond-resolution Bolocam SZ imaging of 45 clusters
and found the $Y_{SZ}$ signal does not scale self-similarly with $M_{2500}$ derived from X-ray observations.
The cause of this mismatch may be due to AGN feedback
regulating star formation and altering the gas properties in cluster cores.
Through accurate lensing mass profiles measuring total mass, free from the assumption of gas HSE, 
and accurate gas mass profiles from Chandra, 
we will determine the radial dependence of $f_{gas} = M_{gas} / M_{tot}$ 
and characterize the efficiency with which AGN input energy into the intracluster medium.

Long term efforts to improve mass scaling relations focus on weak lensing analyses,
for example from Euclid and LSST \citep{Grandis18}.
However, strong lensing analyses will also contribute.
Simulations have shown that joint strong + weak lensing analyses 
yield significantly smaller biases (\~2\%) and uncertainties (\~20\%) 
in virial mass estimates than either weak or strong lensing alone \citep{Meneghetti10}. 
A joint strong plus weak lensing analysis of 20 CLASH clusters based on \HST\ and Subaru observations
\citep{Umetsu16}
shows that the $M_{2500}$ masses can be determined with a fractional total uncertainty of 25\% per cluster at
$<$$M_{2500}$$>$ $= 3.6 \times 10^{14} M_{\odot}$ (or $<$$M_{500}$$>$ $=10^{15} M_{\odot}$), 
accounting for the dominant contributions from intrinsic profile variations \citep{Gruen15}.
Thus a similar analysis of all 41 RELICS clusters would yield an overall mass calibration accuracy of about 4\%
from $r_{2500}$ out to the virial radii, 
providing a legacy mass-profile database for critical tests for models of background cosmology and structure formation.

\subsection{Dark Matter Constraints}
\label{sec:DM}

Self-interacting dark matter (SIDM) particles with cross sections of $\sigma_{\rm DM} / m \sim 0.1 - 0.6 ~ {\rm cm}^2 / {\rm g}$
have been invoked to explain several observational inconsistencies with collisionless cold dark matter (CDM) theory,
including galaxy cluster density profiles, ``missing'' Milky Way satellites, and the ``cusp-core'' problem
(see \citealt{RobertsonA18} and references therein).
Baryonic processes can explain these ``problems'' with varying degrees of success. 
Merging galaxy clusters yield the most robust constraints on SIDM. 
As two clusters collide, galaxies pass straight through the collision, leaving the stripped cluster gas lagging behind. 
We expect dark matter to also pass straight through unless the particles have a significant self-interacting cross section. 
Joint \HST\ lensing plus Chandra X-ray analysis of the Bullet Cluster \citep{Clowe06, Bradac06} constrained 
$\sigma_{\rm DM} / m < 1.25 ~ {\rm cm}^2 / {\rm g}$ (68\% confidence; \citealt{Randall08}).
More recently, \cite{Harvey15} improved this constraint to
$\sigma_{\rm DM} / m < 0.47 ~ {\rm cm}^2 / {\rm g}$ (95\% confidence)
by jointly analyzing \HST\ + Chandra imaging of 30 cluster mergers \citepeg{Bradac08, Dawson12}. 
However, \cite{Wittman18} subsequently analyzed more data (including more \HST\ filters)
on these same 30 clusters and claimed the constraint should be more relaxed:
$\sigma_{\rm DM} / m \lesssim 2 ~ {\rm cm}^2 / {\rm g}$.
RELICS includes 7 cluster mergers analyzed by these papers
plus another 17 confirmed plus 9 possible mergers not in their sample. 
Of the confirmed mergers, all but one has existing Chandra imaging. 
Joint analyses of the lens models plus X-ray data of these
massive cluster mergers will help constrain the
SIDM parameter space toward the astrophysically and theoretically interesting limit of $0.1 ~ {\rm cm}^2 / {\rm g}$.

\subsection{Supernovae}
\label{sec:SN}

RELICS observed each cluster in two epochs separated by 30--60 days to identify supernovae (SNe) and other transient phenomena.
This cadence was designed to catch $z \sim 1 - 2$ SNe near peak brightness in epoch 2.
The RELICS proposal included 20 orbits for follow-up imaging to obtain light curves for the more interesting supernovae.
Ultimately, RELICS discovered 11 supernovae (\S\ref{sec:SNresults}).

Our primary goal was to significantly contribute to the numbers of known distant and lensed supernovae 
discovered in previous surveys.
CLASH and CANDELS discovered both Type Ia supernovae (SNIa) and Core Collapse supernovae (CCSN) out to $z = 2.5$. 
The observed SNIa rates suggest that Type Ia progenitors are primarily double white dwarf systems \citep{Graur14}
that do not explode quickly after formation \citep{Rodney14}. 
High-z CCSN rates from these programs reinforced measurements of the cosmic star formation rate history 
and put constraints on the initial mass function for CCSN progenitors \citep{Strolger15}.

Lensed Type Ia supernovae from CLASH and the Frontier Fields
provided the first empirical tests verifying the accuracy of lens model magnification estimates \citep{Patel14, Nordin14, Rodney15}.
The appearance of ``SN Refsdal'' \citep{Kelly15} was the first strongly-lensed SN 
observed as multiple images with measurable time delays \citep{Rodney16}.  
These time delays can be used to test lens models \citep{Treu16, Kelly16} 
or as a cosmological distance measurement \citep{Vega-Ferrero18, Grillo18}.  
Strong lensing clusters have also revealed other faint transient phenomena, 
including extreme magnifications of individual stars or stellar eruptions in galaxies at 
$z = 1.49$ \citep{Kelly18},
$z = 1.01$ \citep{Rodney18}, and
$z = 0.94$ \citep{Chen19, Kaurov19}.
%
%
RELICS observations will also provide a baseline for the longer term (decade) monitoring required to detect higher-redshift supernovae
and other lensed transients, perhaps including individual Population III stars \citep{Windhorst18}.




\begin{figure}
\centerline{
\includegraphics[width = \columnwidth]{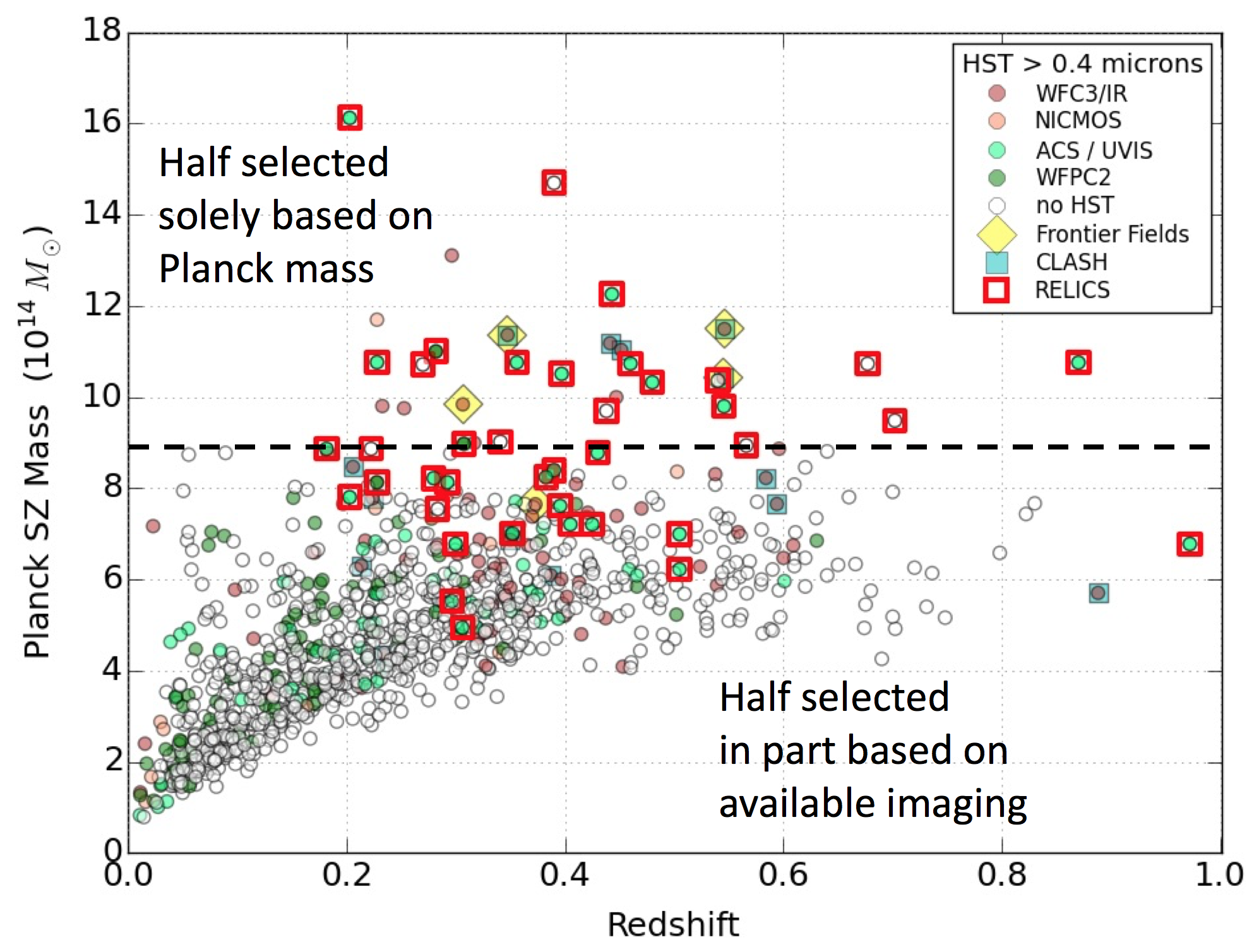}
}
\caption{
\label{fig:MzHST}
RELICS clusters marked as red squares on a plot of mass vs.~redshift for the 1,094 clusters in the Planck PSZ2 catalog.
Each cluster is plotted as a circle color-coded according to existing \HST\ imaging prior to RELICS, 
prioritizing WFC3/IR (red) followed by NICMOS (salmon), ACS and/or UVIS (aqua), and finally WFPC2 (green).
Clusters without \HST\ imaging prior to RELICS are colored white.
Frontier Fields and CLASH clusters are plotted as filled yellow diamonds and blue squares, respectively.
A dashed line at $8.7 \times 10^{14} M_\odot$ separates the 21 clusters selected solely by mass
from the other 20 selected based on various criteria, including existing imaging revealing lensing strength.
Note two RELICS clusters are not plotted here, as they were not included in the Planck PSZ2 catalog.
}
\end{figure}

\begin{figure}
\centerline{
\includegraphics[width = \columnwidth]{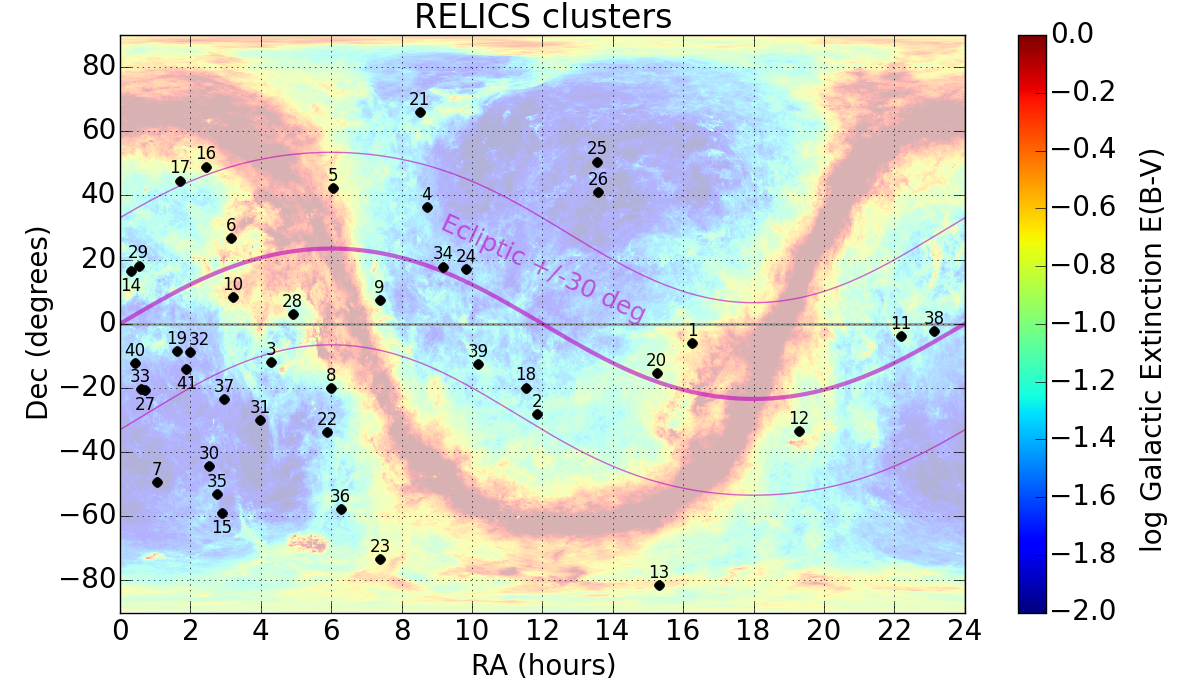}
}
\caption{
\label{fig:skymap}
Distribution on the sky of the 41 RELICS clusters relative to the Galactic and ecliptic planes.
The background is color coded to show the Galactic extinction map from \cite{Schlegel98}.
Numbers correspond to ordered Planck mass as given in Table \ref{tab:clusters}.
}
\end{figure}

\begin{deluxetable*}{rlcccrrc}
\tablecaption{\label{tab:clusters}RELICS Clusters}
\tablehead{
\colhead{}&
\colhead{}&
\colhead{R.A.\supb}&
\colhead{Decl.\supb}&
\colhead{}&
\multicolumn{2}{c}{Planck Mass $M_{500}$}\\[-6pt]
\colhead{Index}&
\colhead{Cluster\supa}&
\colhead{(J2000)}&
\colhead{(J2000)}&
\colhead{Redshift}&
\colhead{Rank\supc}&
\colhead{(\tentothe{14}\Msun)}&
\colhead{E(B-V)}
}
\startdata
1&
Abell 2163 NE&
16:15:48.3&
$-$06:07:36.7&
0.203&
1&
$16.12^{+0.30}_{-0.29}$&
0.2972\\
&Abell 2163 SW&16:15:42.6&$-$06:09:22.1\\
2&
PLCK G287.0+32.9&
11:50:50.8&
$-$28:04:52.2&
0.390\supd&
2&
$14.69^{+0.39}_{-0.42}$&
0.0813\\
3&
MACSJ0417.5-1154&
04:17:33.7&
$-$11:54:22.6&
0.443&
4&
$12.25^{+0.53}_{-0.55}$&
0.0320\\
4&
Abell 697&
08:42:58.9&
$+$36:21:51.1&
0.282&
10&
$11.00^{+0.37}_{-0.37}$&
0.0333\\
5&
RXS J060313.4+4212 N&
06:03:12.2&
$+$42:15:24.7&
0.228&
11&
$10.76^{+0.45}_{-0.43}$&
0.1933\\
&RXS J060313.4+4212 S&06:03:25.6&+42:09:53.6\\
6&
MACS J0308.9+2645&
03:08:55.7&
$+$26:45:36.8&
0.356&
12&
$10.76^{+0.63}_{-0.65}$&
0.1776\\
7&
ACT-CL J0102-49151 NW&
01:02:53.1&
$-$49:14:52.8&
0.870&
13&
$10.75^{+0.48}_{-0.47}$&
0.0086\\
&ACT-CL J0102-49151 SE
&01:03:00.0&$-$49:16:22.2\\
8&
RXC J0600.1-2007&
06:00:09.8&
$-$20:08:08.9&
0.460&
14&
$10.73^{+0.51}_{-0.54}$&
0.0433\\
9&
PSZ2 G209.79+10.23&
07:22:23.0&
$+$07:24:30.0&
0.677&
15&
$10.73^{+0.63}_{-0.66}$&
0.0375\\
10&
PLCK G171.9-40.7&
03:12:56.9&
$+$08:22:19.2&
0.270&
16&
$10.71^{+0.49}_{-0.50}$&
0.4477\\
11&
RXC J2211.7-0350&
22:11:45.9&
$-$03:49:44.7&
0.397&
17&
$10.50^{+0.50}_{-0.49}$&
0.0832\\
12&
PLCK G004.5-19.5&
19:17:04.5&
$-$33:31:28.5&
0.540&
19&
$10.36^{+0.68}_{-0.72}$&
0.0790\\
13&
PLCK G308.3-20.2&
15:18:49.9&
$-$81:30:33.6&
0.480&
20&
$10.32^{+0.57}_{-0.58}$&
0.2348\\
14&
RXC J0018.5+1626&
00:18:32.6&
$+$16:26:08.4&
0.546&
24&
$9.79^{+0.53}_{-0.53}$&
0.0501\\
15&
SPT-CL J0254-5857&
02:54:16.0&
$-$58:57:11.0&
0.438&
26&
$9.69^{+0.37}_{-0.38}$&
0.0183\\
16&
PSZ2 G138.61-10.84&
02:27:06.6&
$+$49:00:29.9&
0.702&
27&
$9.48^{+0.67}_{-0.53}$&
0.1830\\
17&
RXC J0142.9+4438&
01:42:55.2&
$+$44:38:04.3&
0.341&
28&
$9.02^{+0.60}_{-0.64}$&
0.0783\\
18&
Abell 1300&
11:31:54.1&
$-$19:55:23.4&
0.308&
30&
$8.97^{+0.46}_{-0.45}$&
0.0440\\
19&
WHL J013719.8-082841&
01:37:25.0&
$-$08:27:25.0&
0.566&
31&
$8.93^{+0.65}_{-0.70}$&
0.0286\\
20&
RXC J1514.9-1523&
15:15:00.7&
$-$15:22:46.7&
0.223&
33&
$8.86^{+0.41}_{-0.46}$&
0.0869\\
21&
Abell 665&
08:30:57.4&
$+$65:50:31.0&
0.182&
34&
$8.86^{+0.32}_{-0.32}$&
0.0400\\
22&
MACS J0553.4-3342&
05:53:23.1&
$-$33:42:29.9&
0.430&
36&
$8.77^{+0.44}_{-0.46}$&
0.0357\\
23&
SMACS J0723.3-7327&
07:23:19.5&
$-$73:27:15.6&
0.390&
43&
$8.39^{+0.33}_{-0.34}$&
0.1893\\
24&
RXC J0949.8+1707&
09:49:50.9&
$+$17:07:15.3&
0.383&
48&
$8.24^{+0.46}_{-0.46}$&
0.0255\\
25&
Abell 1758a NW&
13:32:39.0&
$+$50:33:41.8&
0.280&
50&
$8.22^{+0.27}_{-0.28}$&
0.0122\\
&Abell 1758a SE&
13:32:53.4&+50:31:31.0
\\
26&
Abell 1763&
13:35:18.9&
$+$40:59:57.2&
0.228&
51&
$8.13^{+0.26}_{-0.27}$&
0.0073\\
27&
Abell 2813&
00:43:25.1&
$-$20:37:14.8&
0.292&
52&
$8.13^{+0.37}_{-0.38}$&
0.0178\\
28&
Abell 520 NE&
04:54:19.0&
$+$02:56:49.0&
0.203&
65&
$7.80^{+0.40}_{-0.41}$&
0.0402\\
&Abell 520 SW&
04:54:04.2&
+02:53:41.9
\\
29&
RXC J0032.1+1808&
00:32:11.0&
$+$18:07:49.0&
0.396&
85&
$7.61^{+0.57}_{-0.63}$&
0.1052\\
30&
RXC J0232.2-4420&
02:32:18.1&
$-$44:20:44.9&
0.284&
91&
$7.54^{+0.33}_{-0.32}$&
0.0165\\
31&
Abell 3192\supe&
03:58:53.1&
$-$29:55:44.8&
0.425&
114&
$7.20^{+0.52}_{-0.50}$&
0.0071\\
32&
MACS J0159.8-0849&
01:59:49.4&
$-$08:50:00.0&
0.405&
115&
$7.20^{+0.61}_{-0.68}$&
0.0207\\
33&
MACS J0035.4-2015&
00:35:27.0&
$-$20:15:40.3&
0.352&
133&
$7.01^{+0.45}_{-0.50}$&
0.0187\\
34&
RXC J0911.1+1746&
09:11:11.4&
$+$17:46:33.5&
0.505&
136&
$6.99^{+0.73}_{-0.79}$&
0.0359\\
35&
Abell S295&
02:45:31.4&
$-$53:02:24.9&
0.300&
156&
$6.78^{+0.37}_{-0.36}$&
0.0445\\
36&
SPT-CL J0615-5746&
06:15:54.2&
$-$57:46:57.9&
0.972&
157&
$6.77^{+0.49}_{-0.54}$&
0.0362\\
37&
MACS J0257.1-2325&
02:57:10.2&
$-$23:26:11.8&
0.505&
227&
$6.22^{+0.70}_{-0.74}$&
0.0251\\
38&
Abell 2537&
23:08:22.2&
$-$02:11:32.4&
0.297&
376&
$5.52^{+0.51}_{-0.51}$&
0.0798\\
39&
MS 1008.1-1224&
10:10:33.6&
$-$12:39:43.0&
0.306&
504&
$4.94^{+0.57}_{-0.60}$&
0.0601\\
40&
MACS J0025.4-1222&
00:25:30.3&
$-$12:22:48.1&
0.586&
$\cdots$&
$\cdots$&
0.0239\\
41&
CL J0152.7-1357&
01:52:42.9&
$-$13:57:31.0&
0.833&
$\cdots$&
$\cdots$&
0.0126
\\
\vspace{-0.1in}
\enddata
\tablenotetext{a}{Cluster name and portion (e.g., NE) if observed by RELICS with two WFC3/IR pointings.}
\tablenotetext{b}{Coordinates of \HST\ WFC3/IR pointings.}
\tablenotetext{c}{Mass rank among all 1,094 clusters in the PSZ2 catalog.}
\tablenotetext{d}{Redshift updated in \cite{Zitrin17}; previously $z = 0.38$.}
\tablenotetext{e}{Observations centered on MACSJ0358.8-2955 ($z = 0.428$) 
include contributions from Abell 3192 ($z = 0.168$).}
\end{deluxetable*}

\begin{deluxetable*}{rp{1.5in}p{5in}}
\tablecaption{\label{tab:clusternotes}RELICS Cluster Notes}
\tablehead{
\colhead{Index}&
\colhead{Cluster}&
\colhead{Notes}
}
\def\arraystretch{0.75}
\startdata
1&
Abell 2163&
Most massive cluster according to both Planck PSZ2 and the X-ray analysis by \cite{Mantz10b} (also see \citealt{Arnaud92})\\
2&
PLCK G287.0+32.9&
Not observed by \Hubble\ or \Spitzer\ prior to RELICS. 
Lensed arcs 80\arcsec\ from the BCG \citep[Figure 18]{Gruen14}.
Discovered in early SZ catalog \citep{Planck11SZ}\\
3&
MACSJ0417.5-1154&
One of the most X-ray luminous clusters discovered by MACS: 2.9\e{45} erg \sinv\ between 0.1--2.4 keV \citep{Ebeling10,Pandge19};
JWST GTO target (CANUCS; PI Willott)
\\
4&
Abell 697&
Highest SZ mass Abell cluster yet to be observed by ACS or WFC3/IR (WFPC2 only);
high mass in X-ray analyses \citep{Mantz10b,Piffaretti11};
more modest in weak lensing \citep{Mahdavi13}
\\
5&
RXS J060313.4+4212
\newline = ``Toothbrush'' &
Radio relic merger shock wave discovered by \cite{vanWeeren12} and studied further \citep{Rajpurohit18}\\
6&
MACS J0308.9+2645&
MACS half-orbit F606W, F814W
\\
7&
ACT-CL J0102-49151
\newline = ``El Gordo'' &
Discovered by \cite{Menanteau12}; excellent lens \citep{Zitrin13ElGordo}; 
weak lensing $M_{200} = 3$\e{15}\Msun\ \citep{Jee14};
ACS $2 \times 2$ mosaic F606W;
JWST GTO target (PI Windhorst)
\\
8&
RXC J0600.1-2007&
MACS half-orbit F814W
\\
9&
PSZ2 G209.79+10.23&
no HST, Chandra, XMM
\\
10&
PLCK G171.9-40.7&
no HST
\\
11&
RXC J2211.7-0350&
MACS half-orbit F606W
\\
12&
PLCK G004.5-19.5&
$z = 1.601$ arc at 30\arcsec\ \citep{Sifon14}
\\
13&
PLCK G308.3-20.2
\newline = SMACSJ1519.1-8130 &
MACS half-orbit F606W
\\
14&
RXC J0018.5+1626
\newline = MS0015.9+1609&
Known strong lens \citep{Zitrin11MACS};
ACS $2 \times 2$ mosaic; 
CCCP $M_{2500} = 6.2 \times 10^{14} M_\odot$ 
4th highest measured of 50 clusters \citep{Hoekstra15};
shallow off target NICMOS
\\
15&
SPT-CL J0254-5857&
One of the most massive clusters discovered by SPT \citep{Williamson11}
\\
16&
PSZ2 G138.61-10.84&
Optical counterpart is 4.5\arcmin\ from PSZ1 coordinates
(2:27:06, +49:05:10) from \cite{Planck16XXXVI}
\\
17&
RXC J0142.9+4438&
CIZA \citep{Kocevski07}
\\
18&
Abell 1300&
``rivals A2744'' \cite{MannEbeling12}; Ebeling WFPC2 F606W
\\
19&
WHL J013719.8-082841 \newline = WHL J24.3324-8.477&
Discovered by \cite{Wen12} based on photometric redshifts in SDSS-III.
Renamed in the PSZ2 catalog with coordinates given in decimal degrees.
\\
20&
RXC J1514.9-1523&
radio halo \citep{Giacintucci11}
\\
21&
Abell 665&
similar in mass and redshift to A1689 (excellent lens); ACS F850LP + grism
\\
22&
MACS J0553.4-3342&
Bullet-like merger \citep{MannEbeling12}; 
very hot and luminous merger \citep{Ebeling17,Pandge17};
big red triple arcs; bright star but probably okay
\\
23&
SMACS J0723.3-7327&
MACS half-orbit F606W, F814W; 2 cycles
\\
24&
RXC J0949.8+1707&
\cite{Wong13} \#7: 7th most powerful lensing line of sight expected based on LRGs identified in SDSS
\\
25&
Abell 1758a&
northern component of merger;
WtG $M_{2500} = 6.1 \times 10^{14} M_\odot$
(\citealt{Applegate14}; as compiled by \citealt{Sereno15})
\\
26&
Abell 1763&
\cite{Rines13} highest mass; \cite{Mantz10b} \#15; WFPC2 good cluster
\\
27&
Abell 2813&
arcs close in; half-orbit F606W (PI Smith)
\\
28&
Abell 520
\newline = ``Train Wreck'' &
ACS $2 \times 2$ mosaic (PI Clowe);
merger like Bullet Cluster but with claimed (and disputed) dark core \citep{Mahdavi07,OkabeUmetsu08,Clowe12,Jee14}
\\
29&
RXC J0032.1+1808&
merger like MACS0717 or MACS0416
\\
30&
RXC J0232.2-4420
\newline = ``RBS-0325'' &
arcs to large radius identified in ground-based imaging
\citep{Kausch10} 
\\
31&
Abell 3192
\newline \& MACSJ0358.8-2955&
\HST\ observations centered on MACS0358 ($z = 0.428$);
includes some contribution from A3192 ($z = 0.168$);
\cite{Mantz10b,Hamilton-Morris12,Hsu13}
\\
32&
MACS J0159.8-0849&
arcs to large radius; not many cluster galaxies; \cite{Wong13} \#59
\\
33&
MACS J0035.4-2015&
arcs to large radius
\\
34&
RXC J0911.1+1746&
MACS
\\
35&
Abell S295
\newline = ACT-CL J0245-5302 &  
Known strong lens \citep{Zitrin11MACS};
ACT SZ detection \citep{Menanteau10a};
\cite{Ruel14,Bayliss16};
ACS observations proprietary before inclusion in RELICS
\\
36&
SPT-CL J0615-5746&
our highest redshift cluster at $z = 0.972$; ACS $2\times 2$ mosaic in F606W and F814W
\\
37&
MACS J0257.1-2325&
Known strong lens \citep{Zitrin11MACS};
Frontier Fields contender; 
F555W, F814W;
WtG $M_{2500} = 5.2 \times 10^{14} M_\odot$
(\citealt{Applegate14}; as compiled by \citealt{Sereno15})
\\
38&
Abell 2537&
Frontier Fields contender;
nice arcs in F606W;
CCCP $M_{2500} = 4.6 \times 10^{14} M_\odot$
\citep{Hoekstra15}
\\
39&
MS 1008.1-1224&
weak lensing and X-ray analyses \citep{Lombardi00,Athreya02,EttoriLombardi03};
ACS $2 \times 2$ mosaic
\\
40&
MACS J0025.4-1222
\newline = ``Baby Bullet'' &
well studied merger similar to the Bullet Cluster \citep{Bradac08,Zitrin11MACS}
\\
41&
CL J0152.7-1357&
z=0.8330 still merging at filaments; well studied
\citep{Jee05,DeMarco05,Umetsu05,Maughan06,Murata15}
%
\enddata
\end{deluxetable*}

\section{Galaxy Clusters}
\label{sec:clus}

RELICS observed the 41 massive galaxy clusters listed in Tables \ref{tab:clusters} and \ref{tab:clusternotes}.
None had existing \Hubble\ infrared imaging (WFC3/IR or NICMOS) prior to RELICS.
We selected 21 of the clusters based on exceptionally high mass estimates from Planck
and the other 20 based on other factors revealing or suggesting exceptional lensing strength.

\cite{Planck15XXVII}
identified 1653 Sunyaev-Zel'dovich (SZ; \citealt{SunyaevZeldovich70}) sources
in their PSZ2 all-sky ($|b| > 15^\circ$) 
catalog\footnote{Planck PSZ2 catalogs are available and described at:
\href{http://irsa.ipac.caltech.edu/data/Planck/release_2/catalogs/}
{http://irsa.ipac.caltech.edu/data/Planck/release\_2/catalogs/}
and
\href{https://wiki.cosmos.esa.int/planckpla2015/index.php/Catalogues}
{https://wiki.cosmos.esa.int/planckpla2015/index.php/Catalogues}}, 
including 1203 confirmed clusters, 1094 of which had measured redshifts and SZ mass estimates.
A few hundred of these were newly discovered clusters, including some of the most massive known.

We queried the \HST\ observations of all the Planck clusters using the 
Mikulski Archive for Space Telescopes 
(MAST)\footnote{\href{http://archive.stsci.edu/hst/search.php}{http://archive.stsci.edu/hst/search.php}}.
Many had existing \HST\ imaging prior to RELICS, but many more did not.

The 34 most massive clusters 
($M_{500} > 8.8 \times 10^{14} M_{\odot}$)
include some already well studied by \Hubble\ and \Spitzer, 
including 4/6 Frontier Fields clusters\footnote{The Frontier Fields cluster Abell 370 has a lower Planck mass 
$M_{500} = 7.7 \times 10^{14} M_{\odot}$
(though weak lensing analysis yields a far greater
$M_{500} = 1.9 \times 10^{15} M_{\odot}$;
\citealt{Umetsu11}).
MACSJ0416.1-2403 is PSZ2 G221.06-44.05, 
but the association was not initially made;
without the redshift, no mass estimate was possible in the PSZ2 catalog.
However, the relatively low Planck signal-to-noise ratio of 5.2 suggests a lower mass.}
and 5/25 from CLASH (with 3 common to both surveys).
But others were less well studied.
Nine of the top 34 had yet to be observed by \Hubble\ or \Spitzer\ at all,
and 12 more had yet to be observed by \Hubble\ in the near-infrared with NICMOS or WFC3.
These 21 clusters comprise half the RELICS sample;
RELICS obtained the first \HST\ infrared imaging of these clusters, and \Spitzer\ imaging as needed.

The remaining 20 RELICS clusters are primarily known strong lenses
based on existing \Hubble\ optical imaging
(or ground-based imaging in the case of RXC0232-44; \citealt{Kausch10}).
Existing \HST\ ACS imaging, where available, reduced our total orbit request.
We also considered other factors in this selection,
including X-ray mass estimates (\citealt{Mantz10b}; MCXC \citealt{Piffaretti11});
weak lensing mass estimates 
(\citealt{Sereno15} compilation including Weighing the Giants \citealt{Applegate14},
\citealt{vonderLinden14a}; \citealt{Umetsu14}; \citealt{Hoekstra15});
SDSS clusters \citep{Wong13,Wen12}; 
and other SZ mass estimates from SPT \citep{Bleem15} and ACT \citep{Hasselfield13}
as well as clusters nearly selected for the 
Frontier Fields\footnote{\href{http://www.stsci.edu/hst/campaigns/frontier-fields/frontier-fields-high-magnification-cluster-candidate-list/}{http://www.stsci.edu/hst/campaigns/frontier-fields/frontier-fields-high-magnification-cluster-candidate-list/}}.

The 41 RELICS clusters generally bear the names of the surveys which discovered them:
\cite{Abell58,Abell89};
the Einstein Observatory Extended Medium-Sensitivity Survey
\citep[MS;][]{Gioia90}; 
the ROSAT X-ray All-Sky Galaxy Cluster Survey \citep[RXC;][]{Ebeling98,Ebeling00};
the ROSAT MAssive Cluster Survey 
\citep[MACS and SMACS;][]{Ebeling01,Ebeling07,Ebeling10,Ebeling13,MannEbeling12,Repp16,ReppEbeling18};
an extended radio source discovered in ROSAT
\citep[RXS;][]{vanWeeren12};
the Wen, Han, \& Liu SDSS-III cluster catalog
\cite[WHL;][]{Wen12};
the South Pole Telescope SZ survey \citep[SPT;][]{Bleem15};
the Atacama Cosmology Telescope SZ survey \citep[ACT;][]{Hasselfield13};
and the Planck all-sky SZ survey \citep{Planck11XI,Planck15XXVII}.
Cl J0152.7--1357 at $z = 0.833$ was discovered by \cite{Ebeling00b} in the Wide Angle ROSAT Pointed Survey (WARPS).
Alternate cluster names include
one also identified in the Rosat Bright Survey \citep[RBS;][]{Kausch10}
and another in Clusters in the Zone of Avoidance \citep[CIZA;][]{Kocevski07}.

The Abell clusters are numbered sequentially based on the original catalog.
Numbers in other cluster names codify coordinates, often in J2000 RA HH:MM and Dec DD:MM.
Planck cluster names, instead, give Galactic coordinates in longitude and latitude.
Table \ref{tab:clusternotes} lists names and any alternate names for each RELICS target.





\section{Observations}
\label{sec:obs}

All data from our large \Hubble\ and \Spitzer\ observing programs had no proprietary period.
We describe each of these programs in turn, 
followed by other subsequent large observing programs of RELICS clusters.






\newcommand{\tp}{$^{[2]}$}
\newcommand{\pp}{$^{[2]}$}
\newcommand{\pppp}{$^{[4]}$}
\newcommand{\ppppp}{$^{[5]}$}


\begin{deluxetable*}{rlccccccccc}
\tablecaption{\label{tab:HSTarchival}Archival \HST\ ACS and WFC3/UVIS imaging of RELICS clusters}
\tablehead{
\colhead{Cluster}&
\colhead{Abbreviated}&
\colhead{F390W}&
\colhead{F435W}&
\colhead{F475W}&
\colhead{F555W}&
\colhead{F606W}&
\colhead{F625W}&
\colhead{F775W}&
\colhead{F814W}&
\colhead{F850LP}\\[-6pt]
%
\colhead{Index}&
\colhead{Name\supa}&
\colhead{(s)}&
\colhead{(s)}&
\colhead{(s)}&
\colhead{(s)}&
\colhead{(s)}&
\colhead{(s)}&
\colhead{(s)}&
\colhead{(s)}&
\colhead{(s)}
}
\startdata
1&
abell2163&
&
4664\tp&
&
&
4667\tp&
&
&
9192\tp&
\\
2&
plckg287+32&
&
&
2160&
&
2320&
&
&
4680&
\\
3&
macs0417-11&
&
&
&
&
7152$^{*}$&
&
&
1910&
\\
4&
abell697&
&
&
&
&
&
\\
5&
rxs0603+42&
5294$^{*}$\pp& 
&
&
&
5068$^{*}$\pp&
&
&
10194\pp&
\\
6&
macs0308+26&
&
&
&
&
1200&
&
&
1440&
\\
7&
act0102-49&
&
&
&
&
7680\pppp&
4688\pp&
5024\pp&
1916&
5032\pp\\
8&
rxc0600-20&
&
&
&
&
&
&
&
1440&
\\
9&
plckg209+10&
&
&
&
&
&
\\
10&
plckg171-40&
&
&
&
&
&
\\
11&
rxc2211-03&
&
&
&
&
1200&
&
&
&
\\
12&
plckg004-19&
&
&
&
&
&
\\
13&
plckg308-20&
&
&
&
&
1200&
&
&
&
\\
14&
rxc0018+16&
&
&
&
4500&
17920\pppp&
&
4623&  
4560&
2540\\
15&
spt0254-58&
&
&
&
&
&
\\
16&
plckg138-10&
&
&
&
&
&
\\
17&
rxc0142+44&
&
&
&
&
&
\\
18&
abell1300&
&
&
&
&
&
\\
19&
whl0137-08&
&
&
&
&
&
\\
20&
rxc1514-15&
&
&
&
&
&
\\
21&
abell665&
&
&
&
&
&
2680\\
22&
macs0553-33&
&
4452&
&
&
2092&
&
&
4572&
\\
23&
smacs0723-73&
&
&
&
&
1200&
&
&
1440&
\\
24&
rxc0949+17&
&
&
&
&
1200&
&
&
1440&
\\
25&
abell1758&
&
5072\pp&
&
&
5088\pp&
&
&
10000\pp&
\\
26&
abell1763&
&
&
&
&
&
&
&
&
\\
27&
abell2813&
&
&
&
&
1200&
&
&
&
\\
28&
abell520&
&
9296\pppp&
&
&
9328\pppp&
&
&
18320\pppp&
\\
29&
rxc0032+18&
&
&
&
&
1200&
&
&
1440&
\\
30&
rxc0232-44&
&
&
&
&
&
&
&
&
\\
31&
abell3192&
&
4500&
&
&
3320&
&
&
4620&
\\
32&
macs0159-08&
&
&
&
&
1200&
&
&
&
\\
33&
macs0035-20&
&
&
&
&
1200&
&
&
1440&
\\
34&
rxc0911+17&
&
&
&
4470&
&
&
&
8825&
\\
35&
abells295&
&
1920&
&
&
1936&
&
&
3944&
\\
36&
spt0615-57&
&
&
&
&
8880\pppp&
&
&
12720\ppppp&
\\
37&
macs0257-23&
&
&
&
4500&
&
&
&
8858&
\\
38&
abell2537&
&
&
&
&
2080&
&
&
&
\\
39&
ms1008-12&
&
&
&
&
17856\pppp&
&
2440&
&
2560\\
40&
macs0025-12&
&
&
&
4140&
&
&
&
4200&
\\
41&
cl0152-13&
&
&
&
&
&
19000\pp&
23452\pp&
&
19000\pp\\
\enddata
\tablenotetext{a}{Abbreviated names provided here are used elsewhere in the text and in data product filenames.}
\tablenotetext{[2],[4]}{~~~~~
Multiple pointings with minimal or no overlap are given in superscript brackets.  
Total exposure times listed should be divided by the numbers [2] or [4] to give the depth at each position.}
\tablenotetext{[5]}{~
$2 \times 2$ mosaic plus one extra central pointing;
dividing that exposure time by 2.5 yields 5088 s within the central pointing.}
\tablenotetext{*}{WFC3/UVIS observations.  (All other archival observations listed are ACS.)}
\end{deluxetable*}










\begin{deluxetable*}{rlccccc}
\tablecaption{\label{tab:HSTobs}RELICS \HST\ Imaging with ACS and WFC3}
\tablehead{
&
&
Cluster
&
&
&
&
Parallel
\\[-2.5pt]
%
\colhead{Cluster}&
\colhead{Abbreviated}&
\colhead{F435W}&
\colhead{F606W}&
\colhead{F814W}&
\colhead{WFC3/IR}&
\colhead{WFC3}
\\[-6pt]
\colhead{Index}&
\colhead{Name}&
\colhead{(orbits)}&
\colhead{(orbits)}&
\colhead{(orbits)}&
\colhead{(orbits)}&
\colhead{(orbits)}
}
\startdata
1&
abell2163&
&
&
&
4\tp
\\
2&
plckg287+32*&
1&
*&
*&
2\\
3&
macs0417-11&
1&
&
&
2
\\
4&
abell697&
1&
1&
1&
2&
3\\
5&
rxs0603+42&
&
&
&
4\tp\\
6&
macs0308+26&
1&
0.5&
0.5&
2\\
7&
act0102-49&
1&
&
&
4\tp\\
8&
rxc0600-20&
1&
1&
1&
2&
3
\\
9&
plckg209+10&
1&
1&
1&
2&
3
\\
10&
plckg171-40&
1&
1&
1&
2&
3
\\
11&
rxc2211-03&
1&
1&
1&
2&
3
\\
12&
plckg004-19&
1&
1&
1&
2&
3
\\
13&
plckg308-20&
1&
1&
1&
2&
3\\
14&
rxc0018+16&
1&
&
&
2\\
15&
spt0254-58&
1&
1&
1&
2&
3\\
16&
plckg138-10&
1&
1&
1&
2&
3\\
17&
rxc0142+44&
1&
1&
1&
2&
3\\
18&
abell1300&
1&
1&
1&
2&
3\\
19&
whl0137-08&
1&
1&
1&
2&
3\\
20&
rxc1514-15&
1&
1&
1&
2&
3\\
21&
abell665&
1&
1&
1&
2&
3\\
22&
macs0553-33&
&
&
&
2
\\
23&
smacs0723-73&
1&
0.5&
0.5&
2+6
\\
24&
rxc0949+17&
1&
0.5&
0.5&
2\\
25&
abell1758&
&
&
&
4\tp\\
26&
abell1763&
1&
1&
1&
2+7&
3\\
27&
abell2813&
1&
1&
1&
2&
3\\
28&
abell520&
&
&
&
4\tp\\
29&
rxc0032+18&
1&
0.5&
0.5&
2\\
30&
rxc0232-44&
1&
1&
1&
2&
3\\
31&
abell3192&
&
&
&
2\\
32&
macs0159-08&
1&
1&
1&
2&
3\\
33&
macs0035-20&
1&
0.5&
0.5&
2\\
34&
rxc0911+17&
1&
&
&
2\\
35&
abells295&
&
&
&
2\\
36&
spt0615-57&
1&
&
&
2\\
37&
macs0257-23&
1&
&
&
2\\
38&
abell2537&
1&
&
1&
2\\
39&
ms1008-12&
1&
&
&
2\\
40&
macs0025-12&
1&
&
&
2
\\
41&
cl0152-13&
1&
&
&
2+5\\
\enddata
\tablenotetext{*}{PLCK G287+32 was also awarded Cycle 23 observations under GO 14165 (PI Seitz).  
In addition to the 4 ACS orbits in F475W, F606W, and F814W listed in Table \ref{tab:HSTarchival}, 
that program also obtained 4 orbits (10447 s) of WFC3/IR F110W imaging.
RELICS observed this cluster in ACS F435W and WFC3/IR F105W, F125W, F140W, and F160W,
relinquishing the 2 awarded orbits of F606W and F814W imaging that would have been duplicated.}
\tablenotetext{[2]}{~~Observations were split between two pointings.}
\tablenotetext{+}{~Indicates WFC3 orbits added for supernova follow-up observations (Table \ref{tab:SNobs}).}
\end{deluxetable*}



\begin{deluxetable*}{llcccccc}
\tablecaption{\label{tab:SNobs}HST WFC3/IR and UVIS Follow-Up Imaging of RELICS Supernovae}
\tablehead{
\colhead{Cluster}&
\colhead{Supernova}&
\colhead{Total Orbits}&
\colhead{F105W}&
\colhead{F125W}&
\colhead{F140W}&
\colhead{F160W}&
\colhead{F350LP}
}
\startdata
abell1763 & Nebra & 7 & 0.3 & 2.7 & & 3.7 & 0.3\\
clj0152-13 & Nimrud & 5 & & 2.5 & & 2.5\\
smacs0723-73 & Yupana & 6 & 2 & 1 & 2 & 1 
\enddata
\tablecomments{The full list of RELICS supernova discoveries is given in Table \ref{tab:SN}.}
\end{deluxetable*}

\begin{deluxetable}{lllccr}
\tablecaption{\label{tab:filters}Nominal Exposure Times and Expected Depths}
\tablewidth{\columnwidth}
\tablehead{
\colhead{}&
\colhead{}&
\colhead{$\lambda$\supa}&
\colhead{Exp.\supb}&
\colhead{Depth\supc}&
\colhead{Sens.\supd}
\\[-6pt]
\colhead{Camera}&
\colhead{Filter}&
\colhead{(\um)}&
\colhead{Time}&
\colhead{(AB)}&
\colhead{(nJy)}
}
\startdata
\HST\ ACS/WFC & F435W & 0.43 & 1952 s & 27.2 & 9\\
\HST\ ACS/WFC & F606W & 0.59 & 1991 s & 27.6 & 7\\
\HST\ ACS/WFC & F814W & 0.81 & 2123 s & 27.1 & 11\\
\HST\ WFC3/IR & F105W & 1.06 & 1361 s & 26.6 & 16\\
\HST\ WFC3/IR & F125W & 1.25 & ~\,686 s & 26.0 & 29\\
\HST\ WFC3/IR & F140W & 1.39 & ~\,686 s & 26.2 & 25\\
\HST\ WFC3/IR & F160W & 1.54 & 1861 s & 26.5 & 19\\
\SST\ IRAC & ch1 & 3.6 & 5 hours & 25.0 & 76\\
\SST\ IRAC & ch2 & 4.5 & 5 hours & 24.6 & 108
\enddata
\tablenotetext{a}{Effective ``pivot'' wavelength \citep{TokunagaVacca05}}
\tablenotetext{b}{Total integration times for Abell 1763; slightly longer integrations were obtained for most other clusters.
Much deeper \SST\ integrations (30 hours) were obtained for 10 RELICS clusters.}
\tablenotetext{c}{5\sig\ point source AB magnitude limit.
\HST: within a 0.4\arcsec\ diameter aperture assuming exposure time is split into 4 integrations.
\SST: assuming medium background with 180 integrations of 100 seconds each.}
\tablenotetext{d}{1\sig\ point source sensitivity (nJy) within the same apertures}
\end{deluxetable}

\begin{deluxetable}{cccccc}
\tablecaption{\label{tab:exp}Nominal HST Exposures for a Cluster and Parallel Field with no Archival Imaging}
\tablehead{
\colhead{Epoch / }&
\colhead{Dither\supa}&
\colhead{Prime}&
\colhead{Exposure}&
\colhead{Parallel}&
\colhead{Exposure}
\\[-6pt]
\colhead{Orbit}&
\colhead{Position}&
\colhead{Filter}&
\colhead{time (s)}&
\colhead{Filter}&
\colhead{time (s)}
}
\startdata
1 / 1 & A & F435W & 370 & F350LP & 454\\
1 / 1 & B & F435W & 667 & F350LP & 701\\
1 / 1 & C & F435W & 667 & F350LP & 701\\
1 / 1 & D & F435W & 371 & F350LP & 496\\
\\
1 / 2 & A & F814W & 516 & F140W & 603\\
1 / 2 & B & F814W & 607 & F105W & 703\\
1 / 2 & C & F814W & 607 & F160W & 703\\
1 / 2 & D & F814W & 516 & F125W & 603\\
\\
1 / 3 & W & F140W & 178\\
1 / 3 & W & F105W & 353\\
1 / 3 & Y & F105W & 353\\
1 / 3 & Y & F140W & 178\\
1 / 3 & Z & F125W & 178\\
1 / 3 & Z & F160W & 503\\
1 / 3 & X & F160W & 503\\
1 / 3 & X & F125W & 153\\
\\
2 / 4 & A & F606W & 516 & F125W & 503\\
2 / 4 & B & F606W & 607 & F160W & 703\\
2 / 4 & C & F606W & 607 & F105W & 703\\
2 / 4 & D & F606W & 516 & F140W & 503\\
\\
2 / 5 & Z & F140W & 178\\
2 / 5 & Z & F105W & 353\\
2 / 5 & X & F105W & 353\\
2 / 5 & X & F140W & 203\\
2 / 5 & W & F125W & 178\\
2 / 5 & W & F160W & 503\\
2 / 5 & Y & F160W & 453\\
2 / 5 & Y & F125W & 230
\enddata
\tablenotetext{a}{Each ACS dither position (ABCD) takes another step across the chip gap.
The WFC3/IR dither positions consist of two close ($\sim 0.8''$) pairs (WX and YZ) 
separated by a larger distance ($\sim 6.5''$) to cover the ``death star''.}
\tablecomments{Each epoch begins with ACS on the prime cluster field and WFC3 in parallel; and ends with WFC3/IR in prime.
Epochs 1 and 2 are separated by about a month, or longer when possible.}
\end{deluxetable}

\subsection{Hubble Imaging}
\label{sec:Hubble}

Of the 41 RELICS clusters, 28 had been observed previously 
by \HST\ with ACS and/or WFC3/UVIS (Table \ref{tab:HSTarchival}).
Our 188-orbit \HST\ Treasury Program (Cycle 23; GO 14096; PI Coe; Deputy PI Bradley) 
obtained additional observations
with ACS and WFC3/IR (Table \ref{tab:HSTobs}).
Five clusters required two WFC3/IR pointings for a total of 46 new WFC3/IR images of strongly lensed fields.
For each field, we observed 2 orbits WFC3/IR split among F105W, F125W, F140W, and F160W.
And for each cluster, we observed 3 orbits ACS split among F435W, F606W, and F814W, minus any archival imaging.
For the 18 RELICS clusters without any existing ACS imaging, we observed the full 3 orbits of ACS and, 
in parallel, 3 orbits of WFC3 on a blank field:
1 orbit WFC3/UVIS F350LP and 2 orbits WFC3/IR F105W, F125W, F140W, and F160W.
The parallels use the same filters as in BoRG[z9-10] \citep{Calvi16,Morishita18},
adding area to our high-z search.

For each cluster, we split the observations into two epochs separated by about a month
to enable searches for supernovae and other transients.
Twenty orbits of RELICS were allocated to follow up such Targets of Opportunity (ToO).
We executed these orbits on three of the nine supernovae discovered by RELICS (Table \ref{tab:SNobs}).

Total integration times for each \HST\ orbit vary with the time available between Earth occultations
and the time to acquire or reacquire guide stars.
Table \ref{tab:filters} gives the total integration times in each filter for Abell 1763,
as a representative example.
For most other clusters, the integration times were slightly longer (by up to 15\%).
Each integration was split into four exposures at different dither positions (pointings).
We dithered across the ACS chip gap and WFC3/IR ``death star'' to fill these gaps in the data.

Table \ref{tab:exp} gives the detailed break down of epochs, orbits, and exposure times 
at each of the four dither positions in each filter for a nominal target with no archival imaging.
In the parallel fields, note that the WFC3/UVIS F350LP imaging consists of four exposures at four dither positions,
but the parallel WFC3/IR imaging consists of only two exposures in each filter (one per epoch).

\begin{deluxetable}{cccccc}
\tablecaption{\label{tab:Spitzer}Spitzer IRAC Imaging Programs}
\tablewidth{\columnwidth}
\tablehead{
\colhead{Program}&
\colhead{PI}&
\colhead{TAC}&
\colhead{hours$^{a}$}&
\colhead{depth$^{b}$}&
\colhead{clusters$^{c}$}
}
\startdata
12005 & Brada\v{c} & GO & 99.9 & 1.2 & 26 \\
12123 & Soifer & DDT & 290 & 5 & 34 \\  
13165 & Brada\v{c} & DDT & 167 & 30 & 3 \\ 
13210 & Brada\v{c} & DDT & 55.5 & 30 & 1 \\ 
14017 & Brada\v{c} & GO & 333.2 & 30 & 6 
\enddata
\tablenotetext{a}{Total hours awarded, including overheads.}
\tablenotetext{b}{Resulting total hours observed in each IRAC filter (ch1, ch2) for each cluster, including previous observations.}
\tablenotetext{c}{Number of RELICS clusters observed.  First two programs observed all clusters as required to reach target depths.
Final three programs observed 10 clusters to achieve 30-hour depth for each.}
\end{deluxetable}

\subsection{Spitzer Imaging}
\label{sec:Spitzer}

Altogether, RELICS has been awarded 945 hours of \Spitzer\ observing time (Table \ref{tab:Spitzer}).
For each cluster, we obtained IRAC imaging as needed in the two warm mission filters,
ch1 (3.6\um) and ch2 (4.5\um).
Combined, these filters span approximately $3 - 5$ \um.
About 100 hours of archival IRAC imaging were also available for 18 of the clusters in these two filters,
most notably from GO 60034 (PI Egami).
Below we summarize the complete \Spitzer\ RELICS datasets, which will be detailed further by Strait et al.~(2019, in preparation).

Our initial 100-hour SRELICS (\Spitzer\ RELICS) program (Cycle 12; GO 12005; PI \Bradac) 
was supplemented by a 290-hour Director's Discretionary Time program (DDT 12123; PI Soifer).
Combined, these programs observed all RELICS clusters (as needed) 
to achieve a total of 5 hours integration time (combining new and archival imaging)
in each of the two IRAC filters.
The five clusters requiring two WFC3/IR pointings also required two IRAC pointings.

Subsequently, based on our analyses of the \HST\ and \Spitzer\ images,
we were awarded three more proposals (PI \Bradac)
to obtain deeper IRAC imaging (30 hours / band)
requiring an additional 556 hours
on the 10 clusters yielding the most high-z candidates at $z \sim 6 - 10$ \citep{Salmon17,Salmon18}.
DDT 13165 observed PLCK G287.0+32.9, PLCK G004.5--19.5, and A1763.
DDT 13210 observed SPTCLJ0615--5746 and the $z \sim 10$ arc discovered by \cite{Salmon18}.
And GO 14017 observed CLJ0152.7--1357, ACTCLJ0102--49151, PLCKG308.3--20.2, RXS J060313.4+421, MS1008.1--1124, and SMACSJ0723.3--7327.

\subsection{Other Large Surveys of RELICS Clusters}
\label{sec:surveys}

In addition to the large \HST\ and \SST\ observing programs,
RELICS has also motivated large surveys with other telescopes including VST, VISTA, \XMM, ALMA, and VLA
(see Table \ref{tab:obs}).
The VLA survey exclusively observes RELICS clusters,
while the other listed programs have most or just some clusters in common with RELICS.
We have also carried out smaller follow-up programs 
with many other observatories including 
Keck MOSFIRE \& DEIMOS, VLT MUSE \& X-SHOOTER, Gemini GMOS, 
Subaru SuprimeCam \& HSC, Magellan MegaCam \& LDSS3, MMT Hectospec, and GMRT.
Spectra from Magellan LDSS3 have already been used in several papers 
(\citealt{Cerny18}; \citealt{Paterno-Mahler18}; Mainali et al.~2019, in preparation).
Also note that while we are advertising these newer surveys here, 
we also emphasize that many RELICS clusters have been observed previously by large programs
carried out with \Hubble, \Spitzer, \Chandra, \XMM, and ground-based observatories.


\begin{deluxetable*}{llllccc}
\tablecaption{\label{tab:obs}Large Surveys Including RELICS Clusters Motivated in part by RELICS}
\tablehead{
\colhead{Name}&
\colhead{PI}&
\colhead{Observatory}&
\colhead{Instrument}&
\colhead{Time}&
\colhead{Wavelength}&
\colhead{RELICS / Total Clusters}
}
\startdata
RELICS\supa & Coe & \Hubble\ & ACS, WFC3 & 115 hours & 0.4 -- 1.7 \um & 41 \\ 
S-RELICS\supb & \Bradac; Soifer & \Spitzer\ & IRAC & 945 hours & 3 -- 5 \um & 34\supf \\
GAME & Mercurio & VST & OmegaCAM & 300 hours & ugri & 9 / 12 \\
GCAV\supc & Nonino & VISTA & VIRCAM & 540 hours & YJKs & 13 / 20 \\
Witnessing...\supd & Arnaud \& Ettori & \XMM & EPIC & 833 hours & 0.15 -- 15 keV & 18 / 118 \\ 
ALCS & Kohno & ALMA & 12-m array & 95 hours & 1.1 mm & 16 / 33 \\
Probing...\supe & van Weeren & VLA & 25-m array & 85 hours & 2 -- 4 GHz & 34\supg
\enddata
\tablecomments{The RELICS clusters were observed by many previous programs, 
including MACS \HST\ snapshot programs and large \Spitzer\ programs (see \S\ref{sec:intro}). 
This table does not include those, instead summarizing the more recent programs inspired in part by the RELICS program.}
\tablenotetext{a}{\href{https://relics.stsci.edu}{https://relics.stsci.edu}}
\tablenotetext{b}{\href{https://irsa.ipac.caltech.edu/data/SPITZER/SRELICS/overview.html}{https://irsa.ipac.caltech.edu/data/SPITZER/SRELICS/overview.html}}
\tablenotetext{c}{\href{http://archive.eso.org/cms/eso-archive-news/first-data-release-from-the-galaxy-clusters-at-vircam-gcav-eso-vista-public-survey.html}{http://archive.eso.org/cms/eso-archive-news/first-data-release-from-the-galaxy-clusters-at-vircam-gcav-eso-vista-public-survey.html}}
\tablenotetext{d}{Witnessing the culmination of structure formation in the Universe}
\tablenotetext{e}{Probing cosmic star formation with the JVLA Lensing Cluster Survey}
\tablenotetext{f}{The remaining 7 RELICS clusters already had archival 5-hour depth, and none of them were targeted for 30-hour depth}
\tablenotetext{g}{All RELICS clusters accessible to VLA}
\end{deluxetable*}

\section{Image Reductions and Catalogs}
\label{sec:reductions}

We have made our \HST\ reduced images and catalogs publicly available via 
MAST\footnote{\href{https://archive.stsci.edu/prepds/relics/}{https://archive.stsci.edu/prepds/relics/}}
at \dataset[doi:10.17909/T9SP45]{https://doi.org/10.17909/T9SP45}
and our \Spitzer\ reduced images available via IRSA\footnote{\href{https://irsa.ipac.caltech.edu/data/SPITZER/SRELICS/overview.html}{https://irsa.ipac.caltech.edu/data/SPITZER/SRELICS/}}.
Below we describe the procedures used to generate these data products.

\subsection{Hubble Image Reductions}
\label{sec:imred}

We reduced all \HST\ images obtained by RELICS 
and all archival \HST\ ACS and WFC3/IR images that overlap with the RELICS WFC3/IR images.
Our \HST\ image reduction procedure is similar to that performed on the Frontier Fields \HST\ images \citep{Lotz16}.
Key differences are that we do not produce ``self-calibrated'' ACS images (due to insufficient numbers of exposures),
nor do we correct for time-variable sky emission 
due to Helium line emission at 1.083\um\ which occasionally affects WFC3/IR F105W images \citep{Brammer14ISR}.

We correct all \HST\ images for bias, dark current, and flat fields with up-to-date reference files.
We manually identified and masked any satellite trails.
ACS images were corrected for CTE (charge transfer efficiency) and bias striping.
The multiple ACS exposures were used to automatically identify and reject cosmic rays.
Each WFC3/IR MULTIACCUM exposure consists of multiple samples, enabling ``up-the-ramp'' cosmic ray rejection.
We also masked bad pixels using updated identifications in WFC3/IR images from GO 14114 (G. Brammer, private communication).

We produced inverse variance maps (IVMs) quantifying the uncertainty in each pixel
before accounting for correlated pixel noise (an additional 10-15\%) and Poisson source noise.
These IVMs were used as weights to drizzle-combine the images obtained in each filter.

For each cluster or parallel field,
we aligned all processed \HST\ images
to two common grids with 0.06\arcsec\ and 0.03\arcsec\ resolution.
We used procedures from DrizzlePac, specifically AstroDrizzle \citep{Gonzaga12}
and as outlined in \cite{Koekemoer02, Koekemoer11} and \cite{Lotz16}.
We corrected the absolute astrometry of our images using the Wide-field Infrared Survey Explorer (WISE) point source catalog \citep{Wright10}.

Finally, we produced automatically scaled color images using Trilogy \citep{Coe12}.
Figures \ref{fig:clus1} -- \ref{fig:clus6} show \HST\ ACS + WFC3 color images of all 46 WFC3/IR cluster fields observed by RELICS.
The ACS color images extend to wider areas, which are not shown here, but are available on MAST.
Also on MAST are color images of the 18 RELICS parallel fields observed with WFC3 UVIS and IR.


\begin{deluxetable*}{lccl}
\tablecaption{\label{tab:sex}SExtractor parameters used in the RELICS \Hubble\ source catalogs}
\tablehead{
\colhead{SExtractor parameter}&
\colhead{\tt acs-wfc3}&
\colhead{\tt wfc3ir}&
\colhead{Description}
}
\startdata
\tt DETECT\_MINAREA & 9 & 9 & Contiguous pixels required above detection threshold\\
\tt DETECT\_THRESH & 1 & 1 & Detection threshold ($\sigma$ above background RMS)\\
\tt BACK\_SIZE & 128 & 64 & Background cell size\\
\tt BACK\_FILTERSIZE & 5 & 3 & Background grid size\\
\tt DEBLEND\_NTHRESH & 32 & 64 & Number of threshold levels\\
\tt DEBLEND\_MINCONT & 0.0015 & 0.0001 & Minimum contrast ratio\\
\tt BACKPHOTO\_TYPE & \tt LOCAL & \tt LOCAL & Method for measuring background\\
\tt BACKPHOTO\_THICK & \tt 24 & \tt 24 & Width of rectangular annulus around each object
\enddata
\end{deluxetable*}

\subsection{Hubble Detection and Photometry Catalogs}
\label{sec:catalogs}

Based on the 0.06\arcsec\ resolution \HST\ images, we produced source catalogs 
using techniques similar to those employed for the public CLASH catalogs \citep{CLASH, Coe13}
and the Frontier Fields analysis presented in \cite{Coe15}.

We ran SExtractor \citep{SExtractor} version 2.8.6 in dual-image mode
to detect objects in each field
and define their isophotal apertures for photometry to be measured in each filter image.
For each field, we produce two source catalogs:
\begin{itemize}
  \item {\tt acs-wfc3} (or {\tt acs-wfc3ir}): based on detections in a weighted stack of all \HST\ images (ACS, WFC3/UVIS, and WFC3/IR),
  optimized to detect most objects
  \item {\tt wfc3ir}: based on detections in a weighted stack WFC3/IR images only
  using a finer background grid and more aggressive deblending,
  optimized to detect smaller high-redshift galaxies
\end{itemize}

The stacked images are weighted sums; 
the weights are the IVMs produced by the drizzling software (\S\ref{sec:imred}).
For SExtractor input, we also produce RMS maps equal to 1 / sqrt(weight).

Table \ref{tab:sex} lists the SExtractor parameters used for the {\tt acs-wfc3} and {\tt wfc3ir} catalogs.
The latter is geared toward detections of small high-redshift galaxies,
so we use a smaller background grid and more aggressive deblending to detect small objects near brighter ones.
The {\tt acs-wfc3} catalogs cover the full (larger) ACS field of view and aim to detect whole objects, 
breaking them apart less often.

For detections in either catalog, 9 contiguous pixels are required at the level of the observed background RMS or higher. 
The {\tt acs-wfc3} ({\tt wfc3ir}) background is calculated in 
5$\times$5 (3$\times$3) grids of cells with 128$\times$128 (64$\times$64) pixels in each cell.
We set the deblending of adjacent objects to 128 (64) levels of 0.0015 (0.0001) minimum contrast.

Each object's detection defines an isophotal aperture,
which SExtractor uses in dual-image mode to measure isophotal photometry in every filter in the aligned images.
Just outside this aperture,
we have SExtractor use a 24-pixel-wide rectangular annulus around each object to estimate and subtract the local background in each filter.
We do not perform aperture corrections, as the PSF FWHMs only vary between $\sim 0.07''$--$0.15''$
and we use relatively large isophotal apertures (see discussion in \citealt{CLASH}, \S5.1).

Finally, we correct all photometry for Galactic extinction using the 
IR dust emission maps of \cite{SchlaflyFinkbeiner11}\footnote{Dust extinctions extracted using 
\href{http://irsa.ipac.caltech.edu/applications/DUST/docs/dustProgramInterface.html}
{http://irsa.ipac.caltech.edu/\\applications/DUST/docs/dustProgramInterface.html}}.
Table \ref{tab:clusters} gives the extinction E(B--V) for each cluster from those maps.
These are multiplied by coefficients $A_\lambda$ 
(Table \ref{tab:extcoeff})
for each filter to determine the extinction in magnitudes.

\begin{deluxetable}{llc}
\tablecaption{\label{tab:extcoeff}Galactic Reddening Extinction Correction Coefficients $A_{\lambda}$ for Each Filter}
\tablewidth{\columnwidth}
\tablehead{
\colhead{Camera}&
\colhead{Filter}&
\colhead{Coefficient}
}
\startdata
WFC3/UVIS & F390W   & 4.514\\
ACS & F435W   & 4.117\\
ACS & F475W   & 3.747\\
ACS & F555W   & 3.242\\
ACS, WFC3/UVIS & F606W   & 2.929\\
ACS & F625W   & 2.671\\
ACS & F775W   & 2.018\\
ACS & F814W   & 1.847\\
ACS & F850LP  & 1.473\\
WFC3/IR & F105W   & 1.015\\
WFC3/IR & F110W   & 0.876\\
WFC3/IR & F125W   & 0.757\\
WFC3/IR & F140W   & 0.609\\
WFC3/IR & F160W   & 0.470
\enddata
\end{deluxetable}

\begin{deluxetable*}{r >{\tt}l l}
\tablecaption{\label{tab:catalogs}\HST\ Source Catalog Content}
\tablehead{
\colhead{column}&
\colhead{parameter}&
\colhead{description}
}
\startdata
1 & id & Object ID number\\
2 & RA & Right Ascension in decimal degrees (J2000)\\
3 & Dec & Declination in decimal degrees (J2000)\\
4 & x & x pixel coordinate\\
5 & y & y pixel coordinate\\
6 & fwhm & Full width at half maximum (arcsec)\\
7 & area & Isophotal aperture area (pixels)\\
8 & stel & SExtractor ``stellarity'' (1 = star; 0 = galaxy)\\
9 & ell & Ellipticity = $1 - B/A$\\
10 & theta & Position angle (CCW wrt x axis; degrees)\\
11 & nf5sig & Number of filters with a 5-sigma detection\\
12 & nfobs & Number of filters observed for this object (in the field of view and without bad pixels)\\
13 & f435w\_mag & F435W isophotal magnitude (99 = non-detection; --99 = unobserved)\\
14 & f435w\_magerr & F435W isophotal magnitude uncertainty (or 1-sigma upper limit for non-detection)\\
15 & f435w\_flux & F435W isophotal flux (e--/s)\\
16 & f435w\_fluxerr & F435W isophotal flux uncertainty (e--/s)\\
17 & f435w\_fluxnJy & F435W isophotal flux (nJy)\\
18 & f435w\_fluxnJyerr & F435W isophotal flux uncertainty (nJy)\\
19 & f435w\_sig & F435W detection significance\\
\nodata & \nodata & (photometry in other filters)\\
62\supa & bright\_mag & Brightest magnitude in any filter\\
63 & bright\_magerr & Brightest magnitude uncertainty\\
64 & zb & BPZ most likely Bayesian photometric redshift\\
65 & zbmin & BPZ lower limit (95\% confidence)\\
66 & zbmax & BPZ upper limit (95\% confidence)\\
67 & tb & BPZ most likely spectral type (1--5 elliptical; 6--7 spiral; 8--11 starburst)\\
68 & odds & P(z) contained within {\tt zb} $\pm~0.04(1+z)$\\ 
69 & chisq & $\chi^2$ poorness of BPZ fit: observed vs.~model fluxes\\
70 & chisq2 & Modified $\chi^2$: model fluxes allowed uncertainties \citep{Coe06}\\
71 & M0 & Magnitude used as BPZ prior: F775W or closest available filter\\
72 & zml & Maximum Likelihood (flat prior) most likely redshift\\
73 & tml & Maximum Likelihood (flat prior) most likely spectral type
\enddata
\tablenotetext{a}{Column numbers will vary depending on the number of \HST\ filters observed in each field.}
\end{deluxetable*}

Table \ref{tab:catalogs} summarizes the output in our \HST\ catalogs available on MAST.
Samples are provided from our Abell 697 IR-detection catalog
for object detection and shape measurement (Table \ref{tab:catsample1}),
photometry (Table \ref{tab:catsample2}),
and Bayesian photometric redshifts (Table \ref{tab:catsample3}), which are discussed below.

As a caution to users, 
we note the source catalogs include image artifacts such as diffraction spikes as well as objects poorly segmented by SExtractor.
Additionally, photometry is complicated in crowded fields, 
especially as brighter cluster members contaminate the light from fainter, more distant objects.

\begin{figure}
\centerline{
\includegraphics[width = \columnwidth]{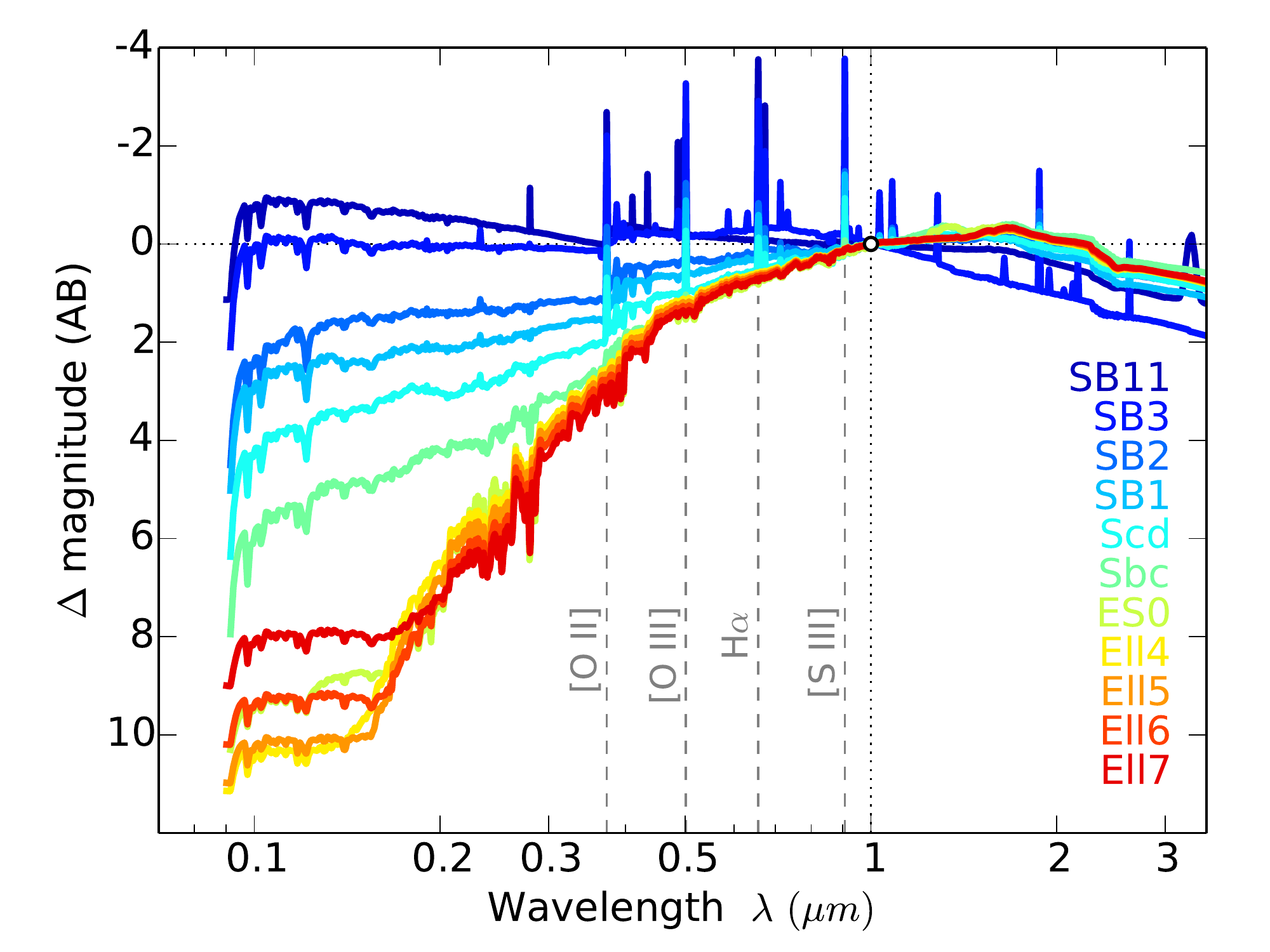}
}
\caption{
\label{fig:seds}
The 11 BPZ template spectral energy distributions (SEDs) used in this work,
consisting of four elliptical galaxies (Ell), one Lenticular (ES0), and four starbursts (SB).
These templates are based on PEGASE \citep{Fioc97},
but recalibrated based on observed photometry and spectroscopic redshifts from FIREWORKS \citep{Wuyts08}.
The starbursts and Scd spiral contain emission lines, four of which are labeled in gray.
All spectra are normalized to the same magnitude at 1 \um.
}
\end{figure}

\begin{figure}
\centerline{
\includegraphics[width = \columnwidth]{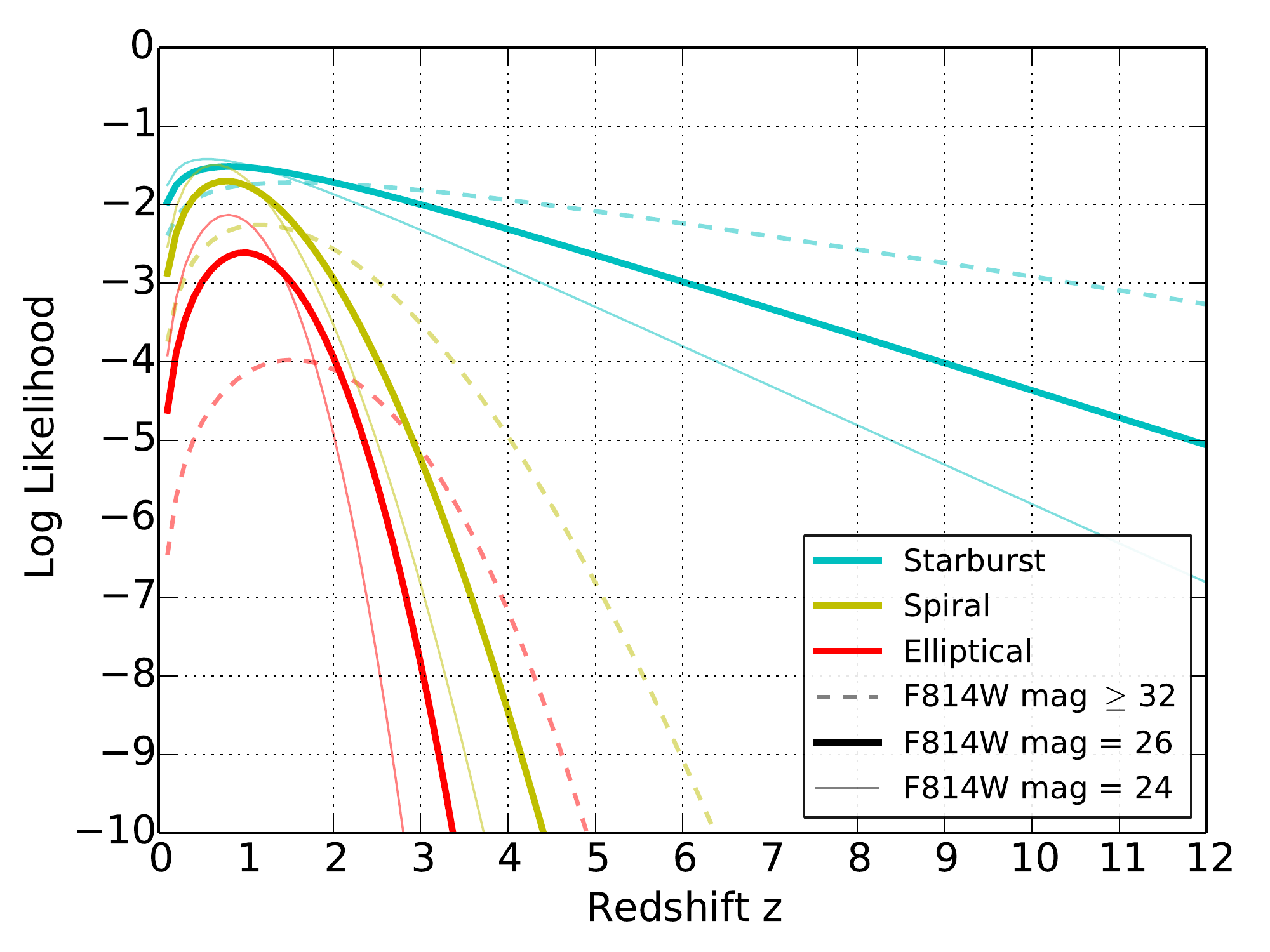}
}
\caption{
\label{fig:prior}
For BPZ, we use the Bayesian redshift prior derived from the HDFN \citep{Benitez00}.
Here we show priors for galaxies of different spectral types 
and with F814W AB magnitudes of 24, 26, and $\geq 32$, including non-detections.
}
\end{figure}

\subsection{Photometric Redshifts}
\label{sec:photoz}

Based on the \Hubble\ photometry, we measured photometric redshifts using
Bayesian Photometric Redshifts (BPZ; \citealt{Benitez00,Coe06})
and Easy and Accurate $z_{phot}$ from Yale (EAZY; \citealt{Brammer08}),
which were two of the top performing methods in controlled tests \citep{Hildebrandt10}.

For BPZ, we used 11 spectral models shown in Figure \ref{fig:seds} and described in \cite{Coe13,Benitez14,Rafelski15}.
Briefly, the model spectral energy distributions (SEDs) are
originally from PEGASE \citep{Fioc97},
but recalibrated, based on observed photometry and spectroscopic redshifts from FIREWORKS \citep{Wuyts08}.
The templates were selected to encompass ranges of metallicities, extinctions, and star formation histories
observed for the vast majority of real galaxies.
We allowed BPZ to interpolate 9 templates between each pair of adjacent templates, yielding 101 templates altogether.
BPZ fit the photometry to a grid of these 101 templates and 1300 redshifts linearly spaced from $z = 0.001$ to $z = 13$.
BPZ tempers those $\chi^2$ results with a Bayesian prior $P(z, T | m)$,
which gives the likelihood of a redshift $z$ and template type $T$ given an observed magnitude $m$ in F814W.
We used the original BPZ prior derived from the HDFN \citep{Benitez00}
and plotted in Figure \ref{fig:prior} for $m = 24$, 26, and $\geq 32$, the latter including F814W non-detections.
We do not attempt to correct magnitudes for lensing magnifications in our initial catalogs 
because those estimates are not available from the start.
We note our prior's dependence on magnitude is often gradual,
but magnification should be accounted for a more accurate $P(z)$ estimate.

We used EAZY \citep{Brammer08} to obtain a second independent set of photometric redshift estimates.
EAZY uses a different template set and allows interpolation between any pair of templates.
The 9 templates used here include 7 from PEGASE, 
plus one very dusty and quiescent galaxy from \cite{Maraston05},
and one extreme emission line galaxy (EELG) from \cite{Erb10}.
We used a flat prior with EAZY as we found the default prior was systematically biased against high-redshift galaxies,
strongly preferring lower redshift EELGs with worse fits to the photometry
(see discussion in \citealt{Salmon17}).

\cite{Salmon17} presents a comparison of the templates and results from BPZ and EAZY for RELICS high-z candidates.

\subsection{Spitzer Image Reductions}

We reduced the \Spitzer\ images using MOPEX \citep{MOPEX} and generated catalogs using T-PHOT \citep{Merlin15}.
The reduced images are available via IRSA.
Currently these include data from programs 12005 and 12123 (totaling 5 hours depth per filter).
Deeper data from the additional three \Spitzer\ programs (Table \ref{tab:Spitzer}) will be included in future releases of reduced images.
More details will be presented by Strait et al.~(2019, in preparation).
Due to the broader \Spitzer\ PSF, extra care is required in obtaining aperture-matched \Hubble\ + \Spitzer\ photometry.
Less careful \Spitzer\ photometry can result in less accurate photometric redshifts \citep{Hildebrandt10}.
We did not use the \Spitzer\ photometry in our initial photometric redshift catalog release.
We did use the \Spitzer\ photometry to vet our $z \sim 10$ candidates, 
and we will use it to study the properties of all our high-redshift candidates.

\begin{deluxetable*}{ccccccrccrcc}
\tablecaption{\label{tab:catsample1}\HST\ Source Catalog: Detection and Shape Measurement}
\tablehead{
\colhead{}&
\colhead{$\alpha_{J2000}$}&
\colhead{$\delta_{J2000}$}&
\colhead{$x$}&
\colhead{$y$}&
\colhead{FWHM}&
\colhead{area}&
\colhead{}&
\colhead{ellipticity}&
\colhead{$\theta$}&
\colhead{}&
\colhead{}
\\[-6pt]
\colhead{ID}&
\colhead{(deg)}&
\colhead{(deg)}&
\colhead{(pixels)}&
\colhead{(pixels)}&
\colhead{($''$)}&
\colhead{(pixels)}&
\colhead{stellarity}&
\colhead{($1-b/a$)}&
\colhead{(deg)}&
\colhead{$N_{f,5\sigma}$}&
\colhead{$N_{f,obs}$}
}
\startdata
1&
130.7320691&
36.3412393&
3212.886&
1118.780&
0.356&
13&
0.35&
0.26&
11.0&
1&
7\\
2&
130.7567476&
36.3856742&
2020.442&
3784.849&
0.320&
474&
0.03&
0.09&
--42.8&
7&
7\\
3&
130.7567488&
36.3851756&
2020.382&
3754.932&
0.754&
125&
0.00&
0.33&
58.4&
7&
7\\
4&
130.7545731&
36.3848770&
2125.475&
3737.004&
0.173&
42&
0.98&
0.06&
--10.0&
5&
7\\
5&
130.7535624&
36.3842826&
2174.292&
3701.339&
0.378&
59&
0.01&
0.21&
80.2&
7&
7
\enddata
\tablecomments{Complete RELICS \HST\ catalogs are available on MAST,
including all parameters described in Table \ref{tab:catalogs}.
This sample of the content is from the Abell 697 IR-detection catalog.}
\end{deluxetable*}

\begin{deluxetable*}{D@{~+/--}DD@{~+/--}DD@{~+/--}Dr}
\tablecaption{\label{tab:catsample2}\HST\ Source Catalog: Photometry in Each Filter}
\tablehead{
\multicolumn4c{Magnitude}&
\multicolumn4c{Flux}&
\multicolumn4c{Flux}&
\\[-3pt]
\multicolumn4c{(AB)}&
\multicolumn4c{($e^-$/s)}&
\multicolumn4c{(nJy)}&
\colhead{S/N}
}
\decimals
\startdata
28.4629 &
0.4579 &
0.0760 &
0.0399 &
14.9573 &
7.8473 &
1.9100\\
24.2190 &
0.0567 &
3.7868 &
0.2031 &
745.4248 &
39.9764 &
18.6500\\
25.5322 &
0.1058 &
1.1298 &
0.1156 &
222.3921 &
22.7527 &
9.7700\\
99.0000 &
28.6848 &
-0.1387 &
0.0619 &
-27.3093 &
12.1918 &
-2.2400\\
25.6583 &
0.0796 &
1.0059 &
0.0765 &
198.0041 &
15.0577 &
13.1500
\enddata
\tablecomments{Continuation of sample provided in Table \ref{tab:catsample1}.
Parameters are described in Table \ref{tab:catalogs}.
The photometry given here is from the F435W filter.
In the full catalog, all filters are provided, followed by the brightest magnitude in any filter for reference.}
\end{deluxetable*}

\begin{deluxetable*}{rrrrrrrrrrrr}
\tablecaption{\label{tab:catsample3}\HST\ Source Catalog: Photometric Redshifts}
\tablehead{
\colhead{$z_{BPZ}$}&
\colhead{$z_{min}$}&
\colhead{$z_{max}$}&
\colhead{$t_{BPZ}$}&
\colhead{ODDS}&
\colhead{$\chi^2$}&
\colhead{$\chi^2_{mod}$}&
\colhead{$M_0$}&
\colhead{$z_{ML}$}&
\colhead{$t_{ML}$}
}
\startdata
2.741&
0.225&
3.470&
9.5&
0.148&
0.940&
1.334&
28.102&
2.730&
9.5\\
0.222&
0.145&
0.255&
4.1&
0.850&
0.933&
0.158&
22.015&
0.190&
4.2\\
1.033&
0.631&
1.115&
9.6&
0.481&
0.163&
0.212&
25.013&
1.040&
9.6\\
4.501&
4.315&
4.698&
8.2&
0.974&
1.292&
1.367&
25.835&
4.520&
8.2\\
0.778&
0.461&
0.961&
10.3&
0.473&
0.112&
0.770&
25.346&
0.780&
10.2
\enddata
\tablecomments{Continuation of sample provided in Table \ref{tab:catsample1}.
Parameters are described in Table \ref{tab:catalogs}.}
\end{deluxetable*}


\section{Results}
\label{sec:results}

To date, RELICS has delivered the following science results
on high-redshift galaxies (\S\ref{sec:highz}),
strong lens modeling (\S\ref{sec:lensing}),
and supernovae (\S\ref{sec:SN}).
Reduced images, catalogs, and lens models are available via MAST and IRSA.
Constraints on cluster masses (\S\ref{sec:clusters})
and the dark matter particle cross section (\S\ref{sec:DM})
require weak lensing data and analyses on a longer timescale.

\subsection{High-Redshift Candidates}
\label{sec:highzresults}

RELICS yielded 321 high-redshift candidates at $z \sim 6 - 8$,
including the brightest known at $z \sim 6$ \citep{Salmon17}.
These galaxies are lensed as brightly as F160W $H \sim 23$,
enabling detailed studies of galaxy properties in the first billion years.
Follow-up study is beginning to match some of these candidates as multiple images \citep{Acebron18a}
and deliver spectroscopic confirmations \citep{Cibirka18}.
We are following up RELICS high-redshift candidates with ground-based telescopes and instruments
including Keck/MOSFIRE, VLT/MUSE, 
Gemini/GMOS, Gemini-S/Flamingos-2,
ALMA, and NOEMA/PdBI.

RELICS also delivered SPT0615-JD, the most distant lensed arc known \citep{Salmon18}.
At $z \sim 10$ and spanning a full $2.5''$ on the sky, 
SPT0615-JD provides by far the most detailed view we have of any galaxy in the first 500 million years.
(The two known $z \sim 11$ galaxies are not spatially resolved,
despite lensing magnification in the case of MACS0647-JD from \citealt{Coe13},
and the relatively high intrinsic luminosity of GN-z11 from \citealt{Oesch16}.)
ALMA observations have been awarded (PI Tamura)
to search for the [\OIII] 88\um\ line in the $z \sim 10$ lensed arc,
continuing the success of that research group at $z \sim 7 - 9$ as noted in \S\ref{sec:highz}
\citep{Inoue16,Tamura18,Hashimoto18a}.
An [\OIII] detection would yield the highest spectroscopic redshift confirmation to date
along with the earliest detection of heavy elements (oxygen).

RELICS \Spitzer\ (SRELICS) imaging was crucial in distinguishing between bonafide $z \sim 10$ candidates
and $z \sim 2$ interlopers.  
\cite{Salmon18} actually identified three $z \sim 10$ candidates based on our \HST\ imaging,
but two turned out to be $z \sim 2$ interlopers based on the \Spitzer\ photometry.
Red $z \sim 2$ galaxies are significantly brighter at 3--5\um\ than bluer $z \sim 10$ galaxies.
%
RELICS \Spitzer\ imaging will also enable us to measure stellar masses for our $>300$ candidates at $z \sim 6 - 8$.

Improved constraints on the $z \sim 9$ luminosity function from RELICS
will require adding simulating lensed galaxies to our images to quantify our detection efficiency 
as a function of magnitude, position, and redshift \citep[e.g.,][]{Livermore16,Carrasco18}.
But based on our current lack of any strong $z \sim 9$ candidates and a single $z \sim 10$ candidate \citep{Salmon17,Salmon18},
our yields appear lower than expected at these redshifts, suggesting support for the accelerated evolution scenario.
Alternatively, more detailed study may shed light on why we have missed galaxies at these redshifts in our searches
and photometric redshift analyses.


At lower redshifts, RELICS is studying compact, low metallicity dwarf galaxies
that are excellent analogs to high-redshift galaxies, but can be studied in greater detail.
Analysis of one RELICS $z=1.645$ galaxy shows that low metallicity stars are driving C\III] emission
with the strongest rest-frame equivalent width (\~20\AA) yet observed at these redshifts,
suggesting a more intense radiation field than assumed by most population synthesis models
(Mainali et al.~2019, in preparation).

\subsection{Lens Modeling}
\label{sec:lensresults}

Strong lens models of RELICS clusters are primarily used to estimate magnifications of our lensed galaxies 
and to correct the surveyed volume for lensing magnification
(\S\ref{sec:lensing}).
We use three lens modeling methods, all of which assume the observed cluster light traces some component of the cluster mass distribution:
Lenstool \citep{Kneib96,JulloKneib09,Johnson14};  
Zitrin LTM, or light-traces-mass \citep{Zitrin09a,Zitrin15};
and GLAFIC \citep{Oguri10}.
The use of multiple methods on an individual cluster
yields a more accurate estimate of systematic uncertainties
than using one method alone,
as shown in analyses of Frontier Fields clusters
\citep{Coe15,Livermore16,Acebron17,Remolina-Gonzalez18}.

RELICS has published strong lens modeling analyses of 14 clusters to date
\citep{Cerny18,Acebron18a,Acebron18b,Cibirka18,Paterno-Mahler18,Mahler18}.
Most of these analyses have revealed lensing strengths on par with Frontier Fields clusters,
quantified in terms of cumulative area with magnification greater than some threshold.
Strong lens models are currently available on MAST for 28 / 41 RELICS clusters,
with the rest to be delivered in time for the \JWST\ GO call for proposals.
Our data products include maps of cluster mass as well as lensing deflection, shear, and magnification.

\cite{Paterno-Mahler18} used all three methods (primarily Lenstool) to model SPT0615-57
and study the $z \sim 10$ candidate discovered by \cite{Salmon18}.
They delivered magnification estimates and explained the current lack of observed counterimages.
\cite{Acebron18b} compared magnification estimates from two methods (Zitrin LTM and Lenstool)
for the many (24) $z \sim 6 - 7$ candidates lensed by CL0152-13.
\cite{Acebron18a} modeled two clusters with the Zitrin LTM method, 
including MACS0308+26, which lenses one of the brightest $z \sim 6$ candidates known.
This paper identified two multiple images of that galaxy lensed to $J \sim 23.2$ and 24.6 AB.
\cite{Mahler18} used spectroscopic redshifts of many arcs from VLT/MUSE \citep{Jauzac19}
to produce a detailed Lenstool model of MACS0417-11, confirming it is a strong lens, 
despite curiously yielding no $z \sim 6 -Ê8$ candidates \citep{Salmon17}.
\cite{Cibirka18} modeled four clusters using the LTM method 
and presented one lensed galaxy showing strong \Lya\ emission with a spectroscopic redshift $z = 5.800$.
%
\cite{Cerny18} modeled five clusters using Lenstool and presented detailed estimates of the statistical and systematic uncertainties.
They present new spectroscopic redshifts and a new method to mitigate modeling uncertainties due to photometric redshifts.



Wide field imaging for weak lensing analyses from Subaru and Magellan is in hand for many RELICS clusters,
as are X-ray data from Chandra and \XMM, along with Planck SZ mass measurements.
Combining these with strong lensing mass measurements from \HST,
as in \cite{Umetsu16} for CLASH clusters,
will improve both precision and accuracy on the overall mass calibration of massive clusters (\S\ref{sec:clusters})
and contribute to constraints on the dark matter particle cross section (\S\ref{sec:DM}).


\begin{deluxetable*}{llllll}
\tablecaption{\label{tab:SN}RELICS Supernovae and HST Follow-Up Imaging}
\tablehead{
\colhead{Cluster}&
\colhead{Supernova\supa}&
\colhead{Abbreviation\supb}&
\colhead{R.A. (J2000)}&
\colhead{Decl. (J2000)}&
\colhead{Notes}
}
\startdata
rxc0949+17 & Eleanor\supc & RLC11Ele & 09:49:47.97 & +17:07:24.9 & cluster member\\
rxc0949+17 & Alexander\supc & RLC11Ale & 09:49:48.07 & +17:07:24.0 & cluster member\\
rxc0949+17 & Antikythera & RLC15Ant & 09:49:48.01 & +17:07:23.0 & cluster member\\
rxc0142+44 & Makapansgat & RLC16Mak & 01:43:16.326 & +44:33:50.65 & parallel field\\
abell1763 & Nebra & RLC16Neb & 13:35:15.13  & +41:00:15.8 & lensed\\
macs0025-12 & Quipu & RLC16Qui & 00:25:31.977 & --12:23:31.80 & cluster member\\
macs0257-23 & Cheomseongdae & RLC16Che & 02:57:07.795  & --23:27:11.69 & lensed or cluster member\\ 
plckg171-40 & Kukulkan & RLC16Kuk & 03:12:59.148 & +08:22:43.60 & cluster member\\
clj0152-13 & Nimrud & RLC16Nim & 01:52:40.352 & --13:57:44.81 & lensed\\
rxc0600-20 & William & RLC17Wil & 06:00:12.227 & --20:07:23.91 & cluster member\\
smacs0723-73 & Yupana & RLC17Yup & 07:23:28.40 & --73:27:03.6 & lensed or cluster member 
\enddata
\tablenotetext{a}{Each supernova was named after a historical relic
with the exceptions of Eleanor and Alexander named after Deputy PI Bradley's children
and William named after PI Coe's newborn son.}
\tablenotetext{b}{Abbreviations include the last two digits of the year of appearance.}
\tablenotetext{c}{Discovered in pre-RELICS imaging based on difference comparison with RELICS imaging.}
\end{deluxetable*}

\subsection{Supernovae}
\label{sec:SNresults}

The RELICS observing strategy (\S\ref{sec:SN})
yielded 11 supernovae, summarized here for the first time in Table \ref{tab:SN}.
Most of these are cluster members, as expected.
The first three supernovae, all found in the RXC0949+17 cluster, were announced in \cite{Rodney15ATel}.
Three of the other RELICS supernovae are lensed, and we obtained follow-up \HST\ imaging of them,
using 20 orbits allocated to RELICS for this purpose (Table \ref{tab:SNobs}).
The most distant, dubbed ``Nebra'', is a $z \sim 2$ candidate type Ia lensed by Abell 1763 \citep{Rodney16ATel}.
If the redshift and type were confirmed, Nebra would be among the most distant SN Ia known.
The current record holder at $z=2.22$ is also lensed \citep{Rubin18}.


\section{Summary}
\label{sec:summary}
\label{sec:discussion}
\label{sec:conclusions}

With RELICS observations complete, 
the 34 most massive Planck clusters ($M_{500} > 8.8 \times 10^{14} M_{\odot}$)
now all have \Hubble\ optical and near-infrared imaging 
as well as \Spitzer\ infrared imaging.
Based on this imaging
we have discovered 322 $z \sim 6 - 10$ candidates,
including the brightest galaxies known at $z \sim 6$ \citep{Salmon17}
and the most distant spatially-resolved lensed arc known, SPT0615-JD at $z \sim 10$ \citep{Salmon18}.
These are among the best and brightest targets for detailed follow-up study from the first billion years after the Big Bang.
Follow-up observations of RELICS fields are currently being carried out
with facilities including Keck, VLT, Subaru, Magellan, MMT, GMRT, and ALMA.

At lower redshifts, RELICS is studying compact, low metallicity dwarf galaxies
that are excellent analogs to high-redshift galaxies, but can be studied in greater detail.
We have also discovered 11 supernovae (Table \ref{tab:SN}).

To date, we have published strong lens modeling analyses of 14 RELICS clusters
\citep{Cerny18,Acebron18a,Acebron18b,Cibirka18,Paterno-Mahler18,Mahler18}.
Many of the clusters modeled so far have proven to be comparable to the strongest lenses known.
By combining our strong lensing analyses of the cluster cores
with weak lensing analyses from ground-based imaging covering the full clusters out to their virial radii,
we can derive robust mass profiles for these clusters.
Our cluster mass measurements will help inform SZ mass scaling relations.

RELICS has proven once again that cluster lensing delivers distant galaxies more efficiently 
than blank field observations.
The efficiency gains are greatest for discoveries of relatively bright galaxies.
We expect these gains to be even greater at higher redshifts. 
Our lens modeling analyses of RELICS clusters will identify which cluster lenses
are truly among the best to use going forward to efficiently search for the first galaxies.

RELICS \Hubble\ reduced images, catalogs, and lens models are available via MAST 
at \dataset[doi:10.17909/T9SP45]{https://doi.org/10.17909/T9SP45}.
RELICS \Spitzer\ reduced images are available via IRSA.

\acknowledgements{



We thank Lindsey Bleem for providing Magellan Megacam LDSS3 images of SPT0254-58 and Abell S295
prior to RELICS to inform our \HST\ observations of these clusters.
We thank Florian Pacaud and Matthias Klein for discussions regarding our Abell S295 \HST\ pointings.
We thank Dale Kocevski for an image of RXC0142+44 obtained with the University of Hawaii 2.2-meter telescope.
And we thank Stella Seitz et al.~for sharing their \HST\ observations of PLCKG287+32 obtained during the same cycle.

We thank the STScI and SSC directors and time allocation committees for enabling these large observing programs.
We are grateful to our \HST\ program coordinator William Januszewski for implementing the RELICS \HST\ observations.
We thank Jennifer Mack for expert mentoring of our \HST\ image reduction gurus R.A.~and S.O.
And we thank Gabriel Brammer for providing an updated WFC3/IR hot pixel mask derived from science observations from GO 14114.

We are grateful to the University of Arizona for hosting our team meeting.

The RELICS \Hubble\ Treasury Program (GO 14096) consists of observations obtained by
the NASA/ESA {\em Hubble Space Telescope (HST)}.
\Hubble\ is operated by the Association of Universities for Research in Astronomy, Inc. (AURA), under NASA contract NAS5-26555.

Data from the NASA/ESA {\em Hubble Space Telescope} presented in this paper 
were obtained from the Mikulski Archive for Space Telescopes (MAST),
operated by the Space Telescope Science Institute (STScI).
STScI is operated by the Association of Universities for Research in Astronomy, Inc.~(AURA)
under NASA contract NAS 5-26555.
The \Hubble\ Advanced Camera for Surveys (ACS) was developed under NASA contract NAS 5-32864.

{\em Spitzer Space Telescope} data presented in this paper
were obtained from the NASA/IPAC Infrared Science Archive (IRSA), 
operated by the Jet Propulsion Laboratory, California Institute of Technology.
Spitzer and IRSA are operated by the Jet Propulsion Laboratory, California Institute of Technology under contract with NASA.

We gratefully acknowledge support from JPL for the \Spitzer\ analysis. 
MB and VS also acknowledge support by NASA
through grant number 17-ADAP17-0247.

Part of this work by WD performed under the auspices of the U.S.~DOE by LLNL under Contract DE-AC52-07NA27344.
KU acknowledges support from the Ministry of Science and Technology of
Taiwan (grant MOST 106-2628-M-001-003-MY3) and from Academia Sinica
(grant AS-IA-107-M01).
OG is supported by an NSF Astronomy and Astrophysics Fellowship under award AST-1602595.
SAR was supported by NASA grant HST-GO-14208 from STScI.
which is operated by Associated Universities for Research in Astronomy, Inc. (AURA), under NASA contract NAS 5-26555.
AM acknowledges the financial support of the Brazilian funding agency FAPESP (Post-doc fellowship - process number 2014/11806-9).
JH was supported by a VILLUM FONDEN Investigator grant (project number 16599).




}



\facility{HST (WFC3, ACS); Spitzer (IRAC); Planck; MAST; IRSA}

\bibliographystyle{aasjournal}
\bibliography{papers}






\begin{figure}
\includegraphics[scale = 0.21]{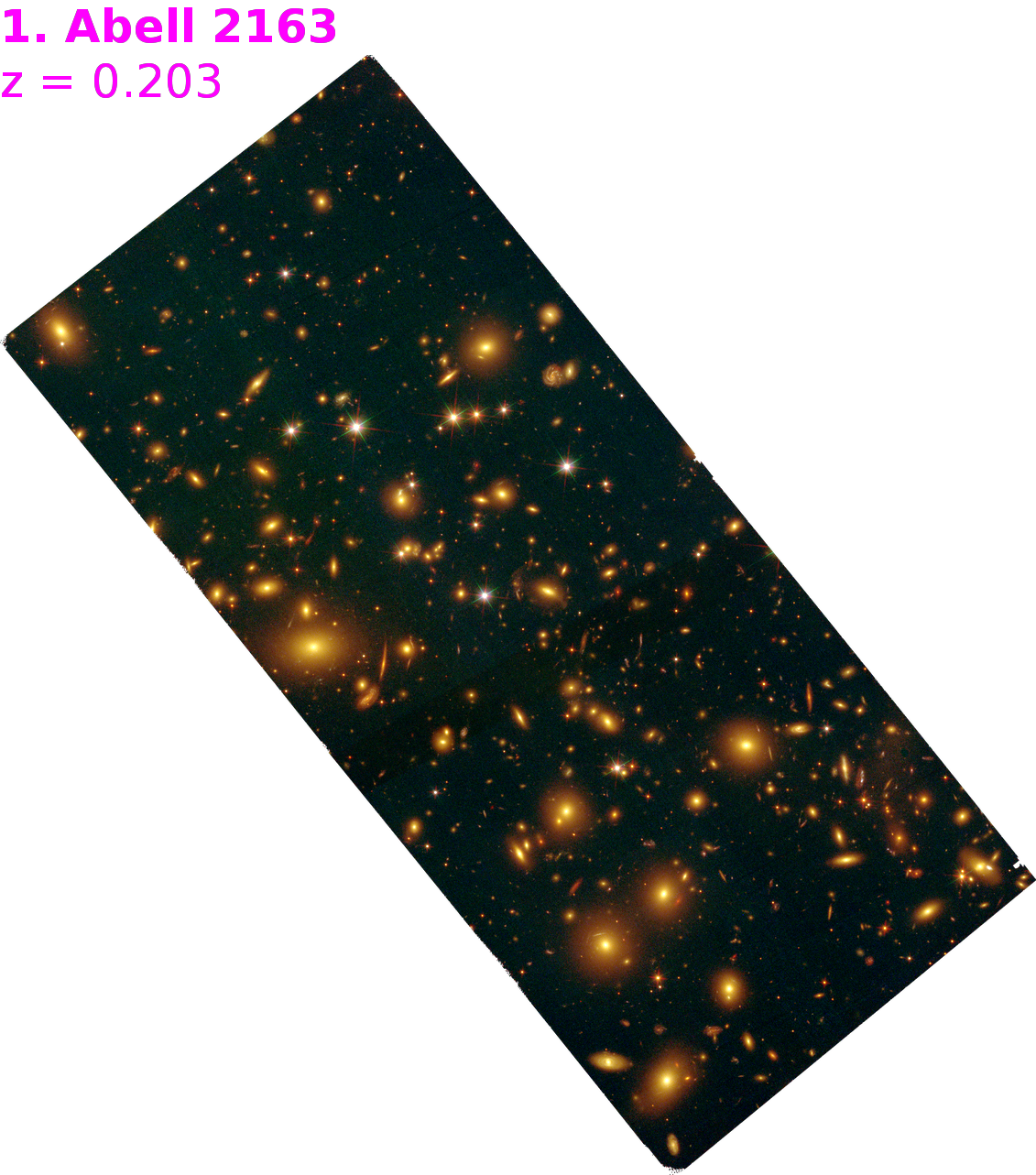}
\includegraphics[scale = 0.21]{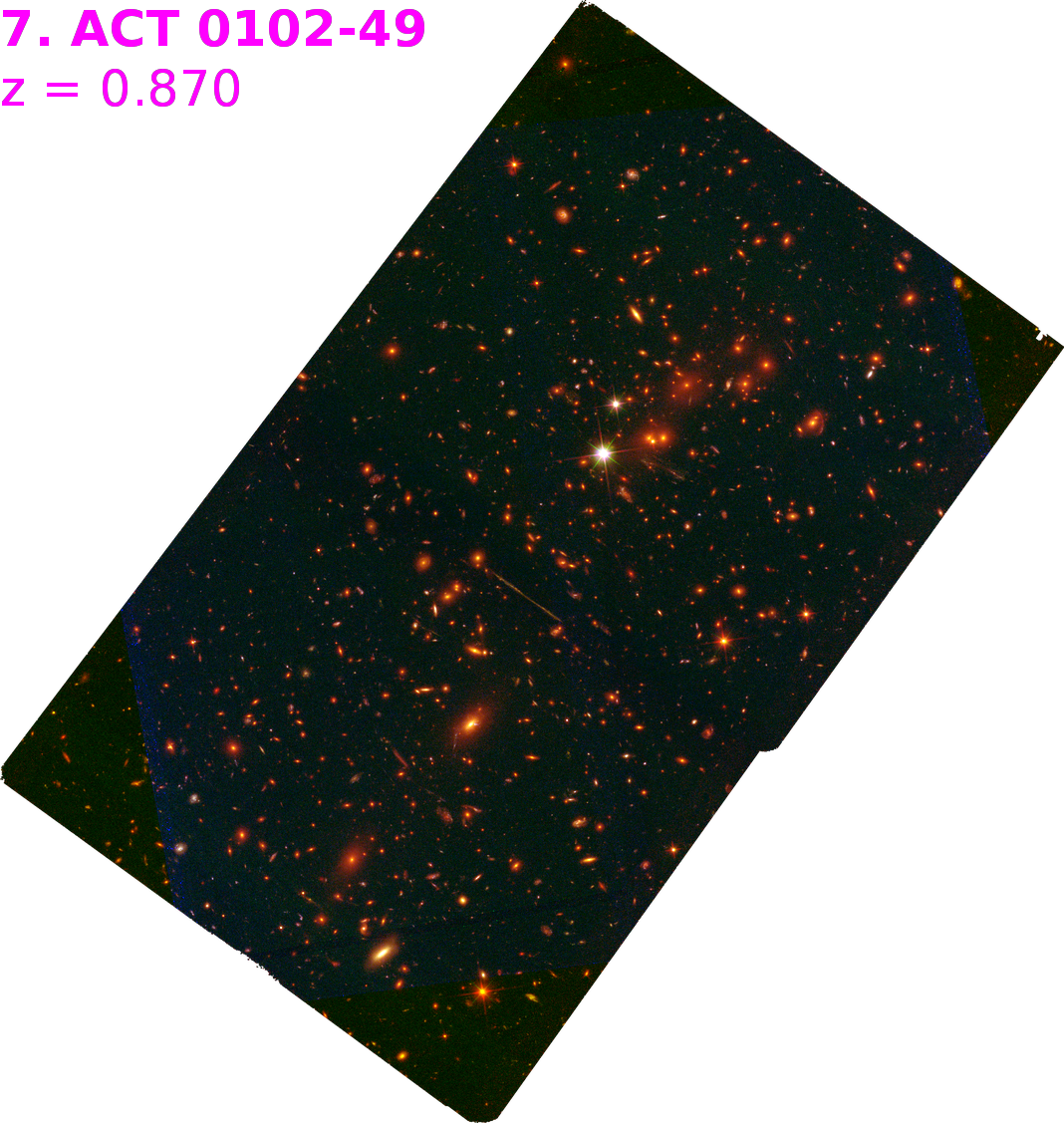}

\caption{
\label{fig:clus1}
\label{fig:clus6}
HST ACS + WFC3/IR observations of RELICS clusters within the WFC3/IR footprints.
ACS imaging extends to wider areas not shown.
All images are shown to the same scale.  North is up; East is left.
Color images produced using Trilogy \citep{Coe12}: blue = F435W; green = F606W + F814W; red = F105W + F125W + F140W + F160W.
Only two examples are shown in the arXiv version of this paper due to file size limitations.
}
\end{figure}


\end{document}